%% file: main.tex
\documentclass[manuscript,acmsmall]{acmart} %anonymous
\AtBeginDocument{%
  \providecommand\BibTeX{{%
    \normalfont B\kern-0.5em{\scshape i\kern-0.25em b}\kern-0.8em\TeX}}}

\setcopyright{acmlicensed}
\copyrightyear{2024}
\acmYear{2024}
\acmConference[Conference acronym 'XX]{Make sure to enter the correct
  conference title from your rights confirmation emai}{June 03--05,
  2024}{Woodstock, NY}

\input{packages}
\input{definitions}

\begin{document}

% \title{User-centric Evaluations of Large Language Models from Real-world Experience}

\title{Understanding User Experience in Large Language Model Interactions}

% Interfacing with Intelligence: A User-Centric Study of LLM Engagement

% Empirical Study on User Intents and Satisfaction for Large Language Models
% User Intents and satisfaction for Large Language Models from Real-world Experience

% User-Centric Evaluations of Large Language Models: Insights from Real-World Interactions

% Beyond Benchmarks: Align Evaluations of Large Language Models with Real-world User Experience

% Unraveling User Experience in Large Language Model Supported Conversational Interfaces

% Beyond Benchmarks: User-centric evaluations of Large Language Models from real-world user experience

% Beyond Benchmarks: Real-world Evaluations into User-centric Language Model Supported Conversational Interfaces

% Beyond Benchmarks: Real-world User Experience in Language Model Supported Conversational Interfaces

% Beyond Benchmarks: Align evaluations of Large Language Model Supported Conversational Interfaces with real-world User Experience

% Beyond Benchmarks: Align evaluations of Large Language Model with real-world User Experience


\author{Jiayin Wang}
\affiliation{%
  \institution{Tsinghua University}
  \city{Beijing}
  \country{China}}
\email{JiayinWangTHU@gmail.com}

\author{Weizhi Ma}
\affiliation{
  \institution{Tsinghua University}
  \city{Beijing}
  \country{China}}
\email{mawz@tsinghua.edu.cn}

\author{Peijie Sun}
\affiliation{
  \institution{Tsinghua University}
  \city{Beijing}
  \country{China}}
\email{sun.hfut@gmail.com}

\author{Min Zhang}
\affiliation{
  \institution{Tsinghua University}
  \city{Beijing}
  \country{China}}
\email{m-z@tsinghua.edu.cn}

\author{Jian-yun Nie}
\affiliation{%
 \institution{University of Montreal}
 \city{Montreal}
 \state{Quebec}
 \country{Canada}}
\email{@iro.edu.cn}

\renewcommand{\shortauthors}{Jiayin Wang, et al.}
\renewcommand{\shorttitle}{User Intent and Satisfaction in LLM Interactions}

\begin{abstract}
In the rapidly evolving landscape of large language models (LLMs), most research has primarily viewed them as independent individuals, focusing on assessing their capabilities through standardized benchmarks and enhancing their general intelligence.
This perspective, however, tends to overlook the vital role of LLMs as user-centric services in human-AI collaboration.
This gap in research becomes increasingly critical as LLMs become more integrated into people's everyday and professional interactions.
% as LLMs grow increasingly integral to both daily and professional settings.
This study addresses the important need to understand user satisfaction with LLMs by exploring four key aspects: comprehending user intents, scrutinizing user experiences, addressing major user concerns about current LLM services, and charting future research paths to bolster human-AI collaborations.

Our study develops a taxonomy of 7 user intents in LLM interactions, grounded in analysis of real-world user interaction logs and human verification. Subsequently, we conduct a user survey to gauge their satisfaction with LLM services, encompassing usage frequency, experiences across intents, and predominant concerns. This survey, compiling 411 anonymous responses, uncovers 11 first-hand insights into the current state of user engagement with LLMs.
% One of them are based on statistical relevance, 7 proposed intents are cluster into 3 groups: subjective use via GUIs, objective use via GUIs and usage through APIs.
Based on this empirical analysis, we pinpoint 6 future research directions prioritizing the user perspective in LLM developments. 
This user-centered approach is essential for crafting LLMs that are not just technologically advanced but also resonate with the intricate realities of human interactions and real-world applications.

\end{abstract}




\maketitle

\input{section/1_introduction}
\input{section/2_related_work}
\input{section/3}

\input{section/4}
\input{section/5}
\input{section/6_Discussion}
\input{section/7}

\clearpage
\bibliographystyle{ACM-Reference-Format}
\bibliography{references}

\clearpage
\balance
\appendix
\input{appendix}

\end{document}

%% file: packages.tex
\usepackage{booktabs} % For formal tables
\usepackage{graphicx}
\usepackage{multirow}
\usepackage[inline]{enumitem}
\usepackage{hhline}
\usepackage{setspace}
\usepackage{makecell}
\usepackage{soul}
\usepackage{xcolor,colortbl}
\usepackage{threeparttable}
\usepackage{balance}
\usepackage{float}
\usepackage{threeparttable}
\usepackage[normalem]{ulem}
\usepackage{wrapfig}
\usepackage{arydshln}
\usepackage{algorithm}
\usepackage{algpseudocode}
\usepackage{amsmath}
\usepackage{graphicx}
\usepackage{subfigure} %需要使用的宏包
\usepackage{bbm}
\usepackage{bm}
\usepackage{diagbox}
 \usepackage{stfloats}
\usepackage{tabularx}
\usepackage[utf8]{inputenc}

%% file: definitions.tex
\useunder{\uline}{\ul}{}

\newcolumntype{L}[1]{>{\raggedright\let\newline\\\arraybackslash\hspace{0pt}}m{#1}}
\newcolumntype{C}[1]{>{\centering\let\newline\\\arraybackslash\hspace{0pt}}m{#1}}
\newcolumntype{R}[1]{>{\raggedleft\let\newline\\\arraybackslash\hspace{0pt}}m{#1}}

\graphicspath{{figures/}}

%% file: section/1_introduction.tex
\section{Introduction}
% \begin{figure}[htbp]
%     \centering
%     \includegraphics[width=3.5in]{Figures/Illustration.png}
%     \caption{User-centric Evaluations}
%     \label{fig:introduction}
% \end{figure}

The rapid evolution of Large Language Models (LLMs) has marked a significant milestone in the field of artificial intelligence and computer technology. 
With their expanding capabilities, LLMs have given rise to new user interfaces, conversational-based service systems, and swiftly amassed a considerable user base.
% Services like ChatGPT, characterized by their fluent comprehension and expression capabilities, along with rich and profound world knowledge and reasoning abilities, are equipped to assist users in accomplishing a wide array of both everyday and professional tasks.
Services like ChatGPT, characterized by their fluent comprehension and expression capabilities, rich and profound world knowledge, and reasoning abilities, are equipped to assist users in accomplishing various everyday and professional tasks~\cite{chang2023survey}.

Despite their increasing involvement in interactions with humans, most research on LLMs has predominantly focused on assessing and improving their abilities as independent intelligence.
For example, there are a lot of benchmarks for evaluating the ability of large language models on specific tasks, such as world knowledge and problem-solving (MMLU \cite{hendrycks2020measuring}), common sense reasoning (HellaSwag \cite{zellers2019hellaswag}, WinoGrande \cite{sakaguchi2021winogrande}), grade school exams (AI2 Reasoning Challenge (ARC) \cite{clark2018think}, GSM-8K \cite{cobbe2021training}), coding (HumanEval \cite{chen2021evaluating}), etc.
These benchmarks give overall scores to create objective ability leaderboards that show how close different models are to general intelligence.
However, this evaluation approach might not align with the actual needs of users who employ LLMs as collaborative tools with both factual and creative intentions.
Furthermore, relying solely on overall performance scores can be misleading; a model with a lower general score might outperform a higher-scoring model in certain situations, depending on specific scenarios and user expectations. This discrepancy highlights the need for user-centric, fine-grained evaluation methods that reflect the practical utility of LLMs in diverse scenarios.
% In contrast, relatively few studies exam LLMs as user-centric interfaces, specifically their effectiveness and capability in meeting user needs within human-computer interaction standpoint.
% Additionally, only the overall performance score for LLMs can be inefficient, as a model with a lower general score may perform better in certain scenarios than one with a higher score, depending on the specific user intent and context.
% \todo{goal: not the evaluate LLM A better than B from benchmarks. understanding user intents and experience in this new interfaces compared to other services}

This paper aims to bridge this research gap as few studies examine LLMs as user-centric interfaces, precisely their effectiveness and capability in meeting user needs from the human-AI collaboration standpoint.
To investigate the practical performances of LLMs in complex real-world scenarios, 
recognizing the diverse intents of users behind the simple inputs is essential for customizing these models to align more closely with user expectations, enabling better user experience and higher efficiency and utilities.

\smallskip
In this paper, we aim to answer the following research questions:
\begin{enumerate}[leftmargin=*,nosep,label=(RQ\arabic*)]
    \item What are the primary \textbf{user intents} for engaging with conversational interfaces powered by large language models~(LLMs)?
    % What motivates users to engage with conversational interfaces powered by large language models?
    \item How do users perceive their \textbf{experience} when interacting with current LLM services in real-world settings?
    \item What \textbf{key concerns} do users have for using large language models?
    % \item What are the gaps in current research in enhancing the user satisfaction with LLMs, and what are \textbf{future directions} for building user-centered generative AI interfaces?
    \item  What are \textbf{future directions} in building user-centered large language models for better human-AI collaboration?
    
    % \item Do existing LLM evaluation benchmarks \textbf{align with} real user experience?
    % \item How can we conduct LLM \textbf{evaluations from a user-centric perspective}, aligning their capabilities with real-world user needs and expectations?
\end{enumerate}
\smallskip

\begin{figure}[htbp]
    \centering
    \includegraphics[width=0.88\linewidth]{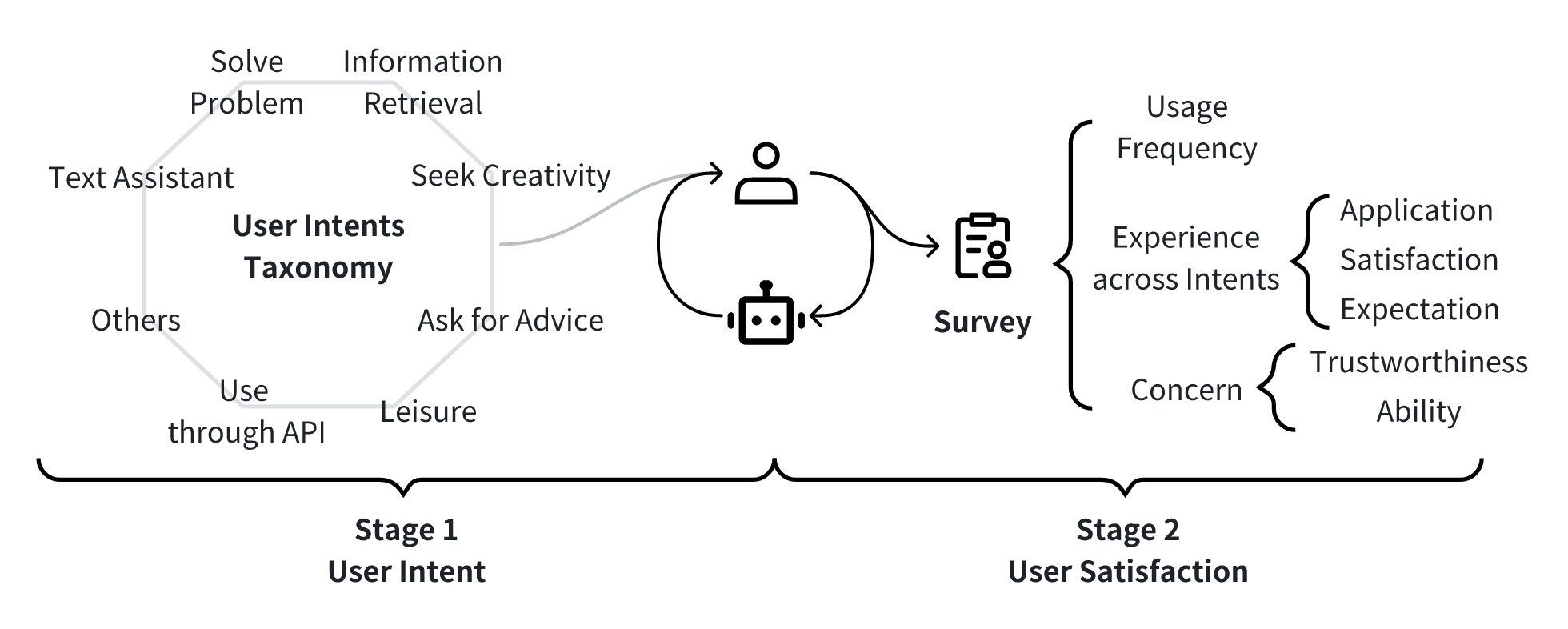} 
    \caption{In this work, we (1) propose the taxonomy of user intents when engaging with large language model interfaces, (2) design and conduct a survey to understand user satisfaction with current LLMs, (3) conclude 11 insightful findings on usage frequency, user experience, and concerns with LLMs, (4) discuss 6 research directions for future human-AI collaboration studies.}
    \label{fig:framework}
\end{figure}

Our study begins by defining a taxonomy of user interaction intents with LLMs based on related analysis and real-world user logs. This classification paves the way for further research into human-LLM interaction dynamics. 
Following the intent categorizations, we conducted an in-depth user survey to assess usage frequency, user experience, and concerns when engaging with large language models. 
With contributions from 411 participants, we obtained valuable firsthand perspectives on user satisfaction with existing LLM interfaces.
Notably, the dataset compiled from this survey will be publicly available alongside the publication of this paper.
We then conduct a thorough empirical analysis of this data, leading to 11 key findings.
% These insights from the user study guide our discussion on crucial future directions regarding large language model research, areas that may not be fully explored in human-AI collaborations.
The insights gleaned from the user study inform our discussion on 6 future research directions in large language model development, highlighting areas in human-AI collaboration that are not fully explored.

% \todo{out dated}
% Our contributions are threefold. First, we categorize real-world user intents for engaging with LLM-based interfaces based on related analysis and real interaction logs. 
% % This categorization provides comprehensive understanding of why users interact with LLMs, revealing a spectrum of intents and purposes. 
% This categorization represents a paradigm shift from task-based evaluations to a focus on usage intents, providing a foundational framework for further user-centric investigations.
% Second, we design and conduct a comprehensive user survey, focused on understanding the user experience across various intents. This survey sheds light on how users perceive and interact with LLMs, offering insights into user satisfaction and concerns.
% Third, we compile and release a dataset of real-world use cases from user surveys. This dataset is tailored for user-centric evaluations, covering a wide range of intents and complexities that reflect actual usage patterns. By doing so, we provide a valuable resource that bridges the gap between current benchmarks and the multifaceted nature of real-world LLM interactions.

\smallskip
To summarize, our main contributions are as follows: We
\begin{enumerate}[leftmargin=*,nosep]
    \item develop a comprehensive taxonomy of user intents for interacting with general large language model interfaces, which is grounded by real-world logs and human verification,
    \item design and implement a survey to gauge user satisfaction with large language models and collect 411 high-quality anonymous user feedback, which will be released along with the publication of this paper,
    % \item and what real-world use cases exemplify these variations
    \item analyze and present 11 insightful findings on the frequency of usage, user experiences across different scenarios, and prevalent concerns regarding large language models,
    \item identify and articulate gaps in existing LLM research with real-world applications, pointing out 6 directions for future studies to enhance user interactions with generative AI interfaces.
    % With suvery and analysis, we determine the extent to which current LLM evaluation benchmarks reflect the coverage and complexity of user interactions in real scenarios and release a dataset with real use cases of the language interfaces for user-centric evaluations.
\end{enumerate}
\smallskip

In conclusion, our study advocates a user-centered approach, emphasizing the need to develop LLMs that are not only technically sophisticated but also genuinely beneficial in human-AI collaborations. This perspective is essential for advancing AI in a way that truly resonates with human needs and real-world utilities.

% As large language models (LLMs) continue to advance at an unprecedented pace, they have garnered significant attention and an active daily user base. These models, which underpin a variety of applications ranging from natural language processing to artificial intelligence, are redefining the landscape of human-computer interaction. However, the traditional methods of evaluating LLMs, primarily focused on measuring their degree of intelligence and capabilities through comprehensive or task-specific scores, may not fully encapsulate their effectiveness in real-world applications. These evaluations often aim to determine the extent to which LLMs approach Artificial General Intelligence, yet they overlook a crucial aspect: the user-centric experience in everyday use.

% In the context of daily interaction, where LLMs serve as an interface for various services, the need for user-centric evaluations becomes paramount. Such evaluations should pivot around the performance of LLMs in specific scenarios, underpinned by a deep understanding of user intents. By recognizing and analyzing these intents, we can more accurately assess how well LLMs cater to the diverse and specific needs of users in real-world settings.

%% file: section/2_related_work.tex
\section{Related Work}
\subsection{User Intent Analysis}
The exploration of user intent in interacting with AI has been a focal point of research.
Extensive work has been conducted on user intents in various information-seeking processes, including web search~\cite{jansen2007determining,yi2009discovering}, product search~\cite{su2018user}, multimedia search~\cite{kofler2016user,tang2011intentsearch}, question-answering and conversational search~\cite{xia2018zero,chen2012understanding,qu2018analyzing,qu2019user}.
For more recent areas like generative AI, there have been only a few works in text-to-image generation systems~\cite{xie2023prompt} and the large language model powered Bing Chat~\cite{shah2023using}, which is also a search-guided product~\cite{kelly2023bing}.

These studies have collectively underscored the significance of understanding user intent and have effectively categorized and used user intents within certain interfaces. Classfication methods include query log analysis~\cite{jansen2006search,bendersky2009analysis} and automated taxonomy generation~\cite{shah2023using}. The proposed well-established intent taxonomy in the search domain categorizes queries as navigational, informational, and transactional~\cite{broder2002taxonomy}. Works~\cite{teevan2008personalize} further discuss personalized search algorithms after understanding user intents.

While user intent is extensively studied in traditional information-seeking services, the advent of generative AI, especially LLMs, introduces new interfaces that are not yet fully explored. 
For instance, conversational search studies categorize intents into original, repeat, clarifying, or follow-up questions~\cite{qu2019user}, focusing mainly on information-seeking behavior, which may not align well with the broader range of interactions in general LLM-powered services, although they share the same multi-round and natural language based characteristics with current LLM services.

The research that most closely aligns with LLM interactions has primarily used closed-source logs from Bing Chat, a search-guided product~\cite{kelly2023bing}, and employed GPT-4 alongside human verification to generate intent taxonomies~\cite{shah2023using}. The taxonomy driven from this work lays the groundwork for our study, as detailed in Section~\ref{sec:intent-validation}.
Our research takes this foundation a step further to interactions within the general LLM-powered environment.

\subsection{Evaluation of Large Language Models}
Both LLM developers release technical reports and research field conduct benchmarks and surveys, most of which are ability-driven evaluations.
Technical reports detail the capabilities of models like GPT-4~\cite{achiam2023gpt}, which showcases proficiency across professional exams, academic benchmarks, and performance in various languages and modalities. Similarly, Mistral~\cite{jiang2023mistral} report highlights performance across tasks like commonsense reasoning, world knowledge, reading comprehension, Math, code, and popular aggregated results, like MMLU~\cite{hendrycks2020measuring}, BBH~\cite{suzgun2022challenging}, and AGI Eval~\cite{zhong2023agieval}.
Additionally, some reports incorporate human-in-the-loop evaluations. The Gemini technical report~\cite{team2023gemini} includes a small section on human preference evaluations, assessing aspects like creativity, instruction-following, and safety through side-by-side blind evaluations by human raters. Llama~\cite{touvron2023llama} also conducts human evaluations focusing on helpfulness and safety. However, these reports often lack details for comparative evaluations between models, and most provide only overall scores without delving into fine-grained performance across different scenarios. 

In the broader research landscape, surveys on LLM evaluations offer a comprehensive overview. One such survey\cite{chang2023survey} categorizes evaluation criteria into areas of natural language processing~\cite{bang2023multitask,liang2022holistic}, robustness/ethics/biases/trustworthiness\cite{wang2023robustness}, social science~\cite{frank2023baby}, natural science and engineering~\cite{bubeck2023sparks}, medical applications~\cite{duong2023analysis}, agent applications~\cite{huang2023language} and other applications like education~\cite{dai2023can}, search and recommendation~\cite{dai2023uncovering}, personality testing~\cite{bodroza2023personality} and specific tasks~\cite{lanzi2023chatgpt}. This classification underscores the multi-dimensional nature of LLM evaluations, but these tasks are often pre-defined
without fully understanding the coverage and difficulty degree of real-world user usage.

% \subsubsection{User-centric Evaluation}

\subsection{Empirical Studies on human-AI Collaborations}

The field of human-AI collaboration has seen extensive research across various domains, including medical AI~\cite{tang2023medical}, legal analysis~\cite{choi2023ai}, tourism~\cite{saydam2022does}, music~\cite{chu2022empirical}, and even areas as unique as romantic love~\cite{song2022can}. Additionally, general guidelines on trust in AI have also been explored~\cite{vereschak2021evaluate}.
These studies delve into critical aspects such as human attitudes towards AI, the development and impact of trust in AI systems, and how AI influences human decision-making~\cite{lai2021towards,mikalef2021artificial}. These diverse research areas collectively contribute to a deeper understanding of the dynamics in human-AI interactions across different fields and contexts. As for the new generative AI interfaces, user experience, such as intents and satisfaction, with general large language models, has been unexplored. Our research hopes to provide a initial step and call for more follow-up work on developing powerful generative AI from a user perspective.

%% file: section/3.tex
\section{Real-world User Intents for Engaging with Large Language Models~(RQ 1)}
Understanding the diverse user intents is essential for tailoring services at a more granular level to suit user demands.
In this context, our work begins by understanding users' primary intents and establishing a taxonomy of user interaction with general-purpose LLMs grounded in related analysis, real-world logs, and further user studies.

\subsection{Taxonomy Development}
\subsubsection{Step 1: Generation Based on Related Literature}
As presented in the previous section, related studies have extensively explored user intents in information systems, with a limited focus on new large language model interfaces. 
Notably, a line of research has proposed user intent taxonomies specifically for LLMs~\cite{shah2023using}, utilizing closed-source conversation logs from Bing Chat, which is a mainly search-oriented product~\cite{kelly2023bing}. We adopt this taxonomy as our starting point, initiating a validation process to assess its applicability to general LLM interfaces.

\subsubsection{Step 2: Validation through Real-World Logs}
\label{sec:intent-validation}
The validation process in related literature~\cite{shah2023using} involved two human coders discussing and deliberating on 30 closed-source Bing Chat conversation segments. 
We proceed the taxonomy further on open-source ChatGPT conversation logs~(ShareGPT~\footnote{ShareGPT is a Chrome extension that lets users share their ChatGPT conversations. We use the data from this source: https://huggingface.co/datasets/shareAI/ShareGPT-Chinese-English-90k.}) to further assess its applicability in general LLM interfaces.
% We utilize the taxonomy of Bing Chat-based literature~\cite{shah2023using}, in which they validate the classification through 30 segments of conversation with two human coders.
% Then we undertake human labeling for real-world ChatGPT conversation logs~(ShareGPT~\footnote{ShareGPT is a Chrome extension that lets user to share their ChatGPT conversations. We use the data from this source: https://huggingface.co/datasets/shareAI/ShareGPT-Chinese-English-90k.}) to further assess its applicability to general LLM interfaces.
We randomly selected 50 English conversation logs and engaged 3 human annotators for validation. 
Initially, annotators independently assessed 10 logs, followed by a phase of discussion and alignment. Subsequently, the annotators collaboratively reviewed and annotated 20 logs in unison. In the final stage, they independently annotated another 20 conversations, culminating in a consensus on the taxonomy.

\begin{figure}[htbp]
    \centering
    \includegraphics[width=\linewidth]{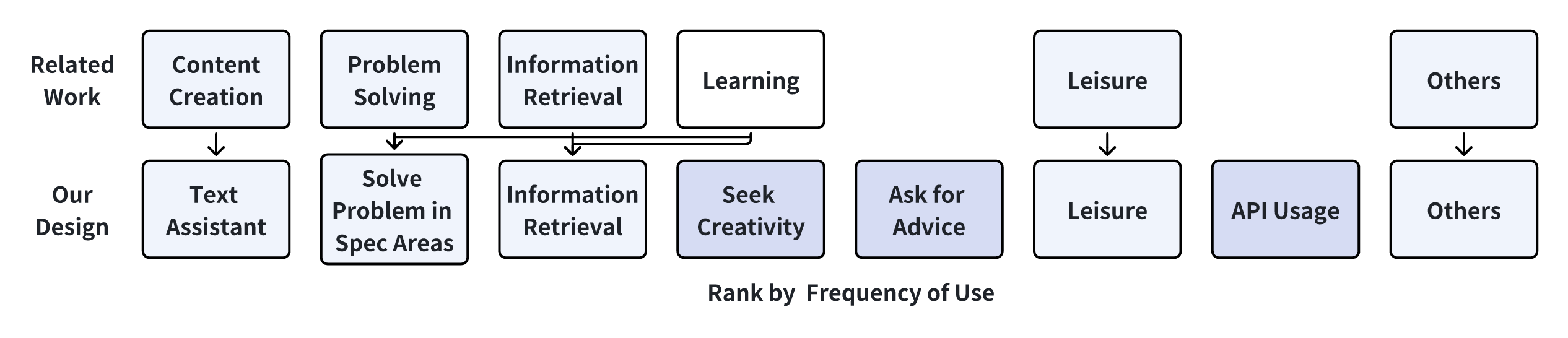}
    \caption{User Intent Taxonomy.}
    \label{fig:intentsDesign}
\end{figure}

This iterative process leads to an enhanced taxonomy~(see Figure~\ref{fig:intentsDesign}), including adding three new intents and consolidating one original category.
``Seek Creativity'' and ``Ask for Advice'' are added, broadening the taxonomy's scope to encompass the general LLM interface usage rather than limiting it to search-related interfaces.
The ``Learning'' category was merged into ``Information Retrieval'' and ``Solve Problem'' depending on whether users acquire direct or inferential information, as ``Learning'' aligns greatly with their information-seeking traits.
Additionally, ``API Usage'' was added to cover interactions through programming interfaces instead of graphical user interfaces. Consequently, we establish a refined user intent taxonomy for general large language model conversational interfaces.

\subsubsection{Step 3: Testing via User Survey}
This taxonomy, especially the newly proposed intents, is further examined through the following user study that involves 411 participants. For details about the users, please see Section~\ref{sec:participant}.

\subsection{Classification result}
\label{subsec:intents}
This section presents our user intent taxonomy when engaging with general large language models. They are introduced in alphabet order~(excluding others):

\smallskip
\begin{enumerate}[leftmargin=*,nosep]
    % \item \textbf{Text Assistant}\\
    % Summarize, translate, revise, generate text, etc.

    % \item \textbf{Solve Problems in Specialized Areas}\\
    %  Seek answers or explanations in the fields of engineering~(especially coding), natural sciences, humanities, social sciences, etc.\\
    %  Address and learn about the profession
    
    % \item \textbf{Information Retrieval}\\
    %  Fast and direct access to factual information
     
    % \item \textbf{Seek Creativity}\\
    % Brainstorming for inspiration, innovative ideas, etc.
    
    % \item \textbf{Ask for Advice}\\
    % Career development, personal counseling, gift recommendation, \\
    % Creating personal schedules, travel plans, shopping lists, etc.
    
    % \item \textbf{Leisure}\\
    % Movie, music, or trip recommendations, gaming, and other entertaining activities
    
    % \item \textbf{Use through API}\\
    % Use through Application Programming Interface instead of user interfaces\\
    % Utilize, test, and explore LLM capabilities, such as evaluating it on various tasks, simulating agents, environments, or datasets, etc.
    
    % \item \textbf{Others}\\
    % Uses that cannot be covered in the above categories
    \item \textbf{Ask for Advice}\\
    Career development, personal counseling, gift recommendation, \\
    Creating personal schedules, travel plans, shopping lists, etc.

    \item \textbf{Information Retrieval}\\
     Fast and direct access to factual information
     
    \item \textbf{Leisure}\\
    Movie, music, or trip recommendations, gaming, and other entertaining activities

     \item \textbf{Seek Creativity}\\
    Brainstorming for inspiration, innovative ideas, etc.

    \item \textbf{Solve Problems in Specialized Areas}\\
     Seek answers, explanations or learn in the fields of engineering~(especially coding), natural sciences, humanities, social sciences, etc.
    
    \item \textbf{Text Assistant}\\
    Summarize, translate, revise, generate text, etc.
    
    \item \textbf{Use through API}\\
    Use through Application Programming Interface instead of graphic user interfaces\\
    Utilize, test, and explore LLM capabilities, such as evaluating it on various tasks, simulating agents, environments, or datasets, etc.
    
    \item \textbf{Others}\\
    Uses that cannot be covered in the above categories
\end{enumerate}

% \smallskip
% This taxonomy is further examined and used through the following user study for fine-grained experience understanding in human-LLM interaction loops. 

%% file: section/4.tex
\section{Design of User Study}
With the established understanding of user intents, we conducted a detailed survey on user satisfaction with current large language models, including usage patterns, experience, and concerns.
% Subsequently, we conducted a detailed user survey to gain insights into usage patterns, user experience, and concerns about large language model interfaces.
% \todo{more explain about the design}
% The feedback from this empirical study provides a deeper understanding of the transformative impact of LLMs on human interaction behaviors and highlights their significance in the evolving landscape of human-computer communication.

\subsection{Questionnaires}
The survey contains 12 questions, with 10 required and 2 optional~(see Appendix~\ref{sec:appendix} for detailed questions),  and takes 5-10 minutes to complete. It is composed of the following five parts:
% \todo{to modified the format}

\begin{enumerate}[leftmargin=*,nosep,align=left]
    \item[Q~1-2] \textbf{Usage Patterns}
    \begin{enumerate}[leftmargin=*,nosep]
    \item[1] \textit{Services Used}
    \item[2] \textit{Frequency of Usage}
    % This aspect is covered by questions about \textit{types of services used} and the \textit{frequency of usage}. This helps to understand the prevalence of LLMs in users' daily lives.
\end{enumerate}
    \item[Q~3-9] \textbf{User Experience across Intents}
    % This part include three major subsection: the intents usage distribution, user satisfaction and expectation.\\
    \begin{enumerate}[leftmargin=*,nosep]
    \item[3-4] \textit{Intents Distribution}\\
    Choose the intents that they have used before.\\
    Opinions about the above intent taxonomy~(optional).
    \item[5] \textit{User Satisfaction across Intents}
    \item[6-8] \textit{User Expectation for Different Answer Types across Intents}\\
    Choose between 3 pairs of answer types: detailed or concise, factual or creative, professional knowledge or common sense.
    \item[9] \textit{User Expectation for Tool Utilization across Intents}\\
    Tools include web browsing, input analysis, personalization, programming, mathematical operations, documentation generation, and multimedia creation.
\end{enumerate}
     \item[Q~10] \textbf{Anchor Question}
\begin{enumerate}[leftmargin=*,nosep]
     \item[10] If the user does not follow the instructions~(select B for this question), this questionnaire would be an invalid response. This helps to control the feedback quality.
\end{enumerate} 
    % \item \textbf{Expected Answer Types and Tool Utilization}\\
     % This indicates a focus on assessing the effectiveness of the language models.
     
    % \item \textbf{Real Use Cases across Intents}\\
    % understand which tools (like web browsing, input analysis, and personalization) contribute to higher-quality responses.
    % Questions about whether the system's responses should be direct or detailed, fact-based or creative, and whether they should rely on professional knowledge or common sense, aim to understand user preferences and expectations in terms of the nature of the responses they receive from the system.
    
    \item[Q~11] \textbf{Major Concerns}
 \begin{enumerate}[leftmargin=*,nosep]
    \item[11] Identify aspects of the system that need optimization, such as hallucinations, long context processing, multi-modal understanding, personalization, privacy, and safety, etc.
    \end{enumerate} 
    \item[Q~12] \textbf{Other Comments}
     \begin{enumerate}[leftmargin=*,nosep]
     \item[12] Comments about the questionnaire or large language model interfaces~(optional).
      \end{enumerate} 
\end{enumerate}

% \smallskip
% Note that we do not record any personal information in the survey, such as age, gender or profession.

\subsection{Participants}
\label{sec:participant}
% The questionnaire was disseminated in both English and Chinese versions. The Chinese version was distributed via WeChat Moments, while the English version was posted on both WeChat Moments and Twitter. The collection period spanned a duration of one month.
% Participants can only fill in the survey once.
% For the Chinese version, participants received 5 RMB (approximately 0.7 USD) for each valid response. For English version, due to payment limitation~(need to tie WeChat account to get the red envelope), we do not give remuneration.

The survey was distributed through WeChat Moments for the Chinese version and via WeChat Moments, a graduate program for international students, and X for the English version. The data collection phase lasted one month, with participants limited to a single submission. 
As we do not record any personal information in the survey, such as age, gender, or profession, the only demographic analysis is the IP distribution automatically recorded by the questionnaire, as detailed in Appendix~\ref{sec:IPs}.
As the survey is spread mainly through social media of graduate students and professors, there might be demographic bias.
This is further discussed in Section~\ref{sec:limitation}.

% For the Chinese version, each valid response earned participants a remuneration of 5 RMB (around 0.7 USD). 
% For English version, due to payment limitation~(need to tie WeChat account to get red envelopes), no remuneration was provided for responses.
% \todo{to be check}

%% file: section/5.tex
\section{Results on User Engagement with LLMs~(RQ 2 and 3)}

\subsection{Feedback Statistic}

We collected 411 feedback, including 297 for the Chinese version and 114 for the English version.
All the feedback passes the anchor question and is treated as valid responses.

In the following subsection, we will report the usage frequency, intent analysis, user satisfaction, expected response type, tool utilization, and major concerns about LLM interfaces. Note that each section is labeled with the corresponding  questionnaire number~(Qx) and has takeaway findings.

\subsection{Usage Frequency~(Q2)}
% \begin{figure}[htbp]
%     \centering
%     \includegraphics[width=5.6in]{Figures/UserIntentsDesign.png}
%     \caption{User Intents}
%     \label{fig:introduction}
% \end{figure}

\begin{figure}[htbp]
    \centering
    \subfigure[Chinese questionnaire]{
        \includegraphics[width=0.475\linewidth]{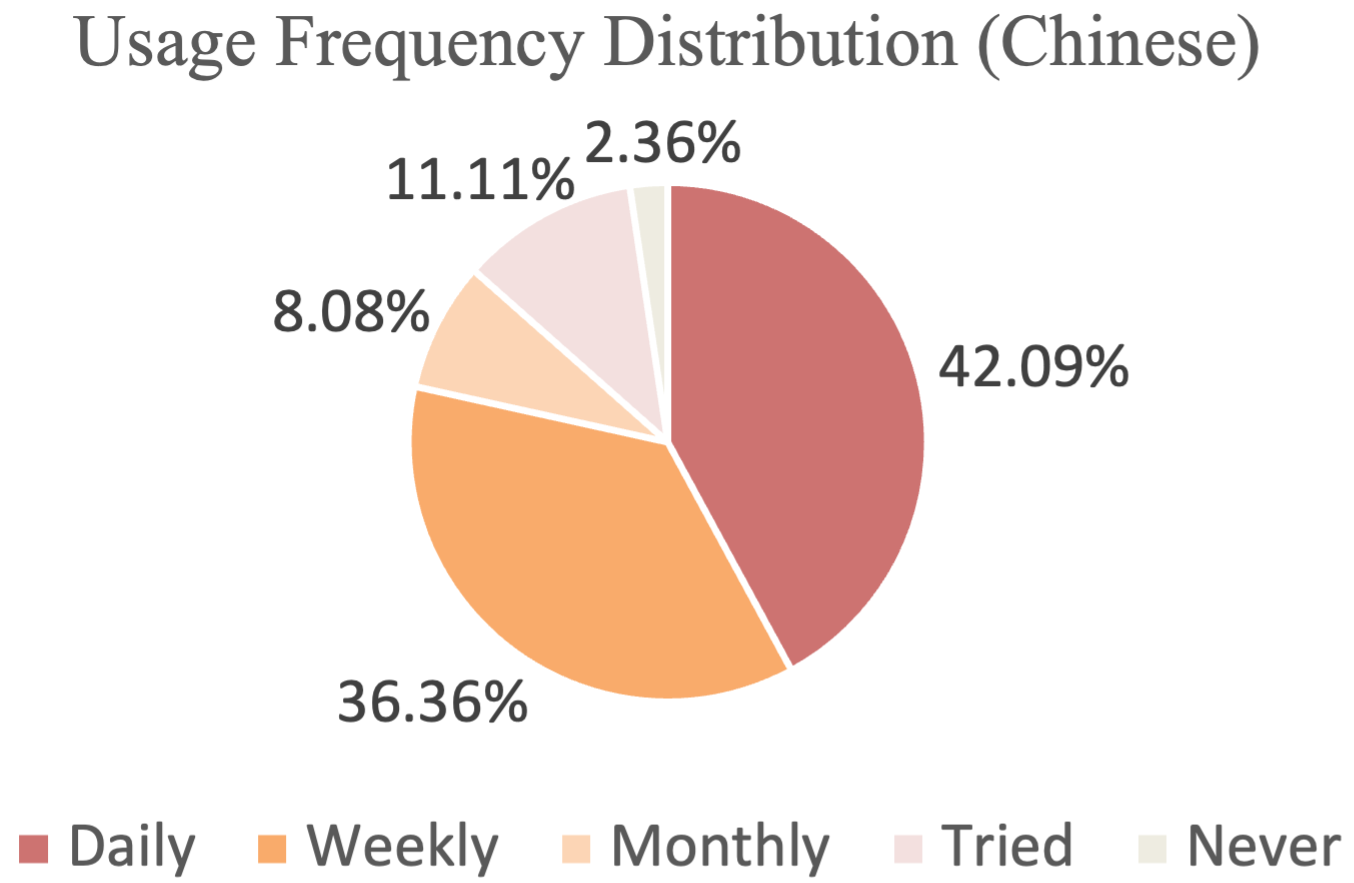}
        \label{fig:Chinese-version}
    }
    \subfigure[English questionnaire]{
	\includegraphics[width=0.475\linewidth]{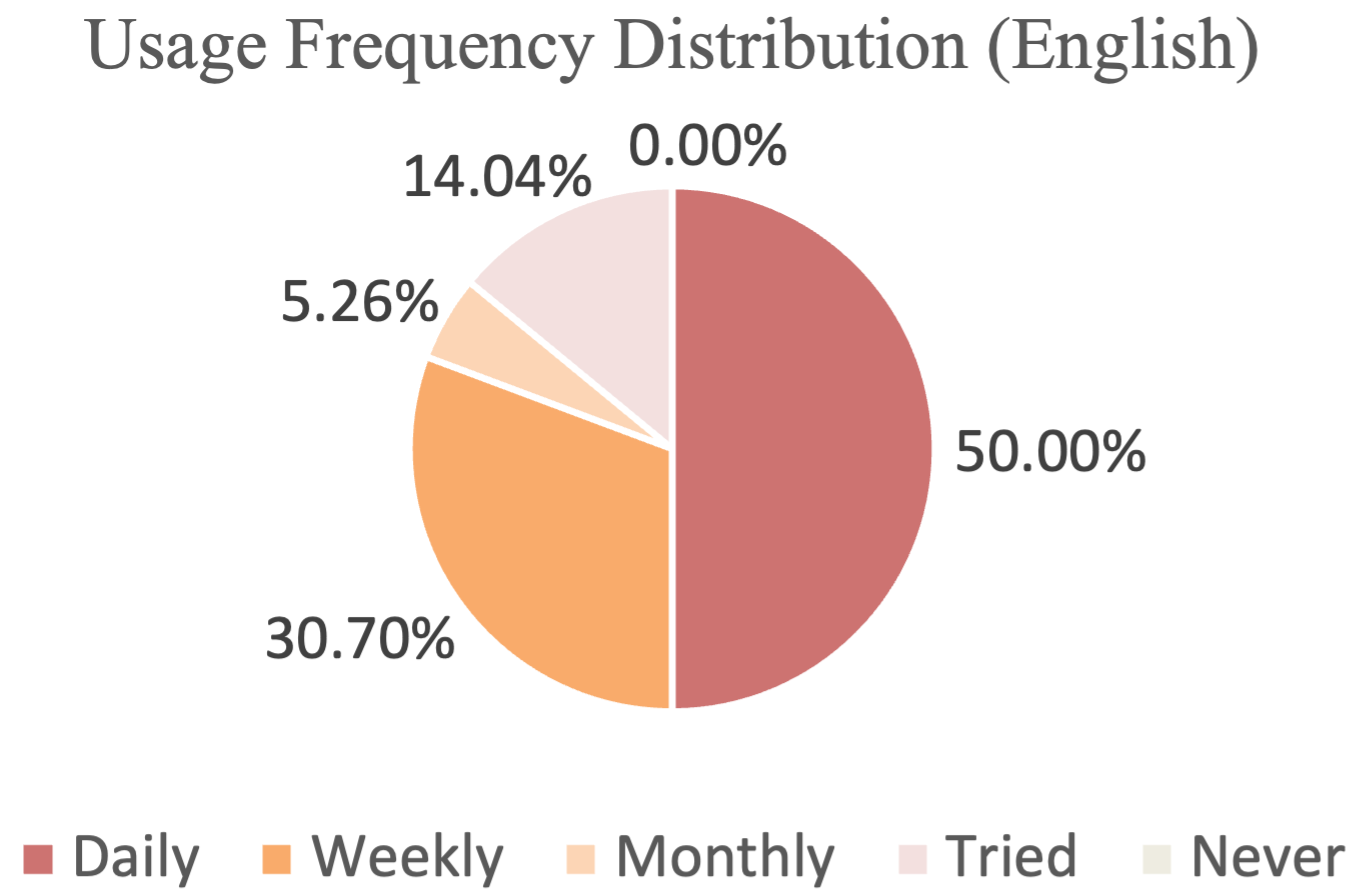}
        \label{fig:English-version}
    }
    \caption{Usage Frequency of the LLM-powered interfaces. Results show that a great number of users interact with large language models on a daily basis.}
    \label{fig:usage-frequency}
\end{figure}

The usage frequency data, as illustrated in Figure~\ref{fig:usage-frequency}, reveals insights into the popularity of large language models. Notably, half of the English and 42.09\% of Chinese respondents report daily use of LLMs, emphasizing their growing importance in everyday activities. Additionally, approximately 80\% of participants from both language groups engage with LLMs at least weekly.
% , underlining their popularity as AI-driven interfaces.
While acknowledging the potential demographic bias as introduced in Section~\ref{sec:participant}, it remains an incontrovertible observation that within this sampled group, large language models are extensively utilized.

% The data also shows that 11.11\% of Chinese and 14.04\% of English respondents have experimented with LLMs, indicating an interest in exploring this new interface, though not necessarily adopting regular usage. 

Compared to a small percentage (2.36\%) of Chinese users who have never used LLMs, the complete adoption among English respondents suggests a potentially higher market penetration in English-speaking demographics. This trend may be influenced by various factors, including perhaps differences in product functionality across languages and the general willingness of populations to embrace new technologies

\medskip
 \fbox{\parbox{0.93\linewidth}{
 % Takeaways:\\
\textbf{\textit{Finding 1}}: Large language model interfaces are used at least weekly by around 80\% of participants.
}
}

\subsection{Intent Analysis~(Q3,4)}

\subsubsection{Usage Distribution}
\label{sec:usage-distribution}
\begin{figure}[htbp]
    \centering
    \includegraphics[width=0.75\linewidth]{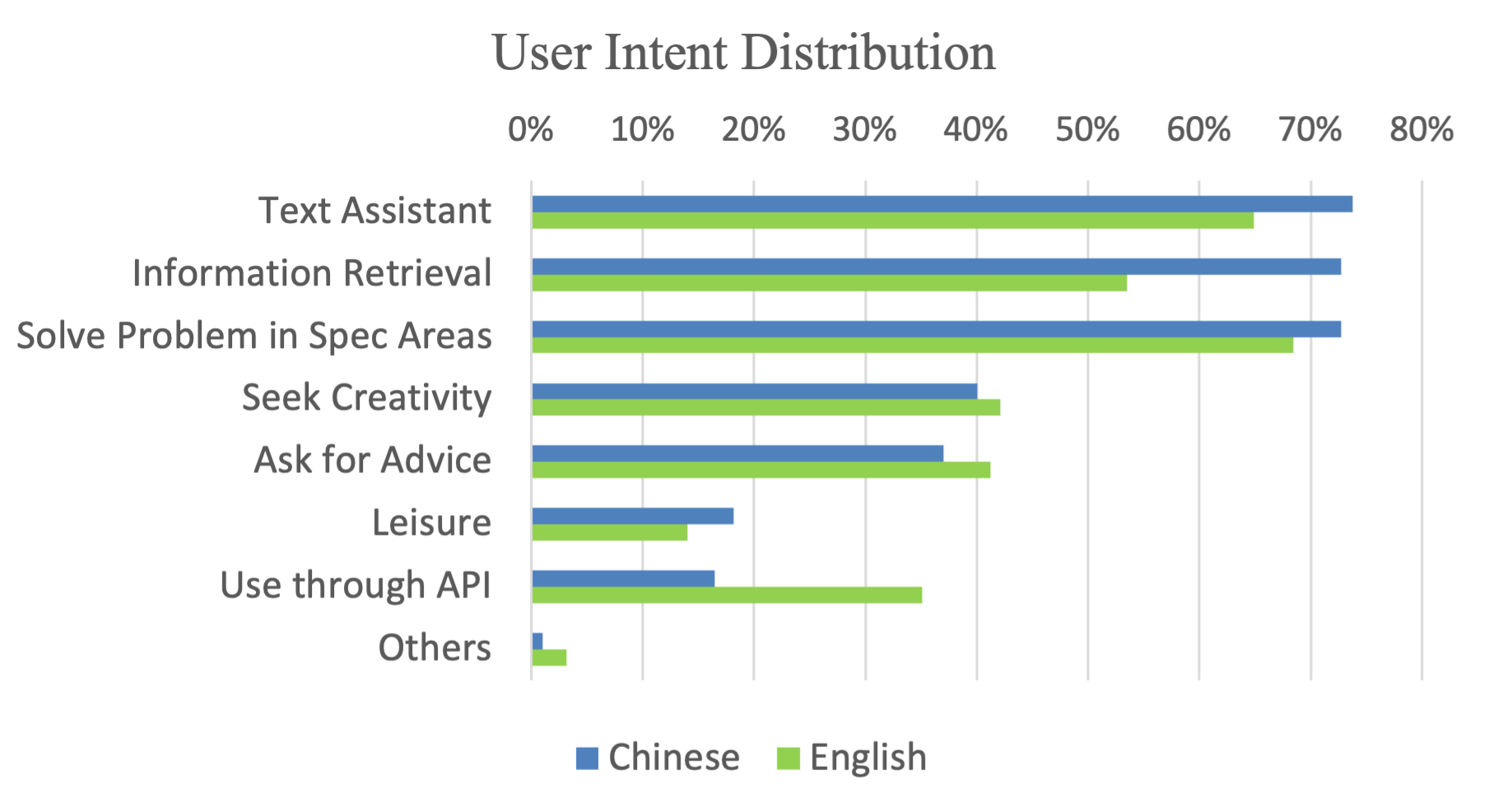}
    \caption{User Intent Distribution: the percentage of users who reported using LLMs under each intent. The intents are ranked from top to bottom according to their frequency of usage in the Chinese questionnaire.}
    \label{fig:intent-distribution}
\end{figure}

Analyzing user intent distribution in the language interface, as shown in Figure~\ref{fig:intent-distribution}, reveals patterns in real-world applications of large language models.

Predominantly, LLMs are used under intents \textbf{Text Assistant}, \textbf{Information Retrieval}, and \textbf{Solve Problem}, indicating a high demand for tasks involving information seeking and processing. This trend likely stems from user familiarity with existing tools such as search engines and high demands for paperwork and professional assistants. This result suggests a lower barrier to adopting these new interfaces as productivity-enhancing tools.

The newly proposed intents, \textbf{Seek Creativity} and \textbf{Ask for Advice}, resonate with about 40\% of users, validating the inclusion of these categories in our intent taxonomy. However, engagement with LLMs for \textbf{Leisure} purposes is considerably lower, possibly due to the models' focus on improving objective metrics like fact-based accuracy, which might overlook elements desired in leisurely interactions, such as serendipity and humor. 
% Interestingly, the \textbf{Use through API} is notably higher among English users, pointing to a greater inclination towards their integration at a technical level in English-speaking communities. 
Notably, English users exhibit a higher propensity for \textbf{Use through API}, suggesting a greater inclination towards embracing new AI tools for scale-up developments.

This discrepancy in using LLMs for subjective versus objective tasks underscores the importance of conducting tailored analyses across different intents. Such fine-grained assessments are crucial for developing LLMs that align closely with human needs and behaviors, fostering better human-AI collaboration. Additionally, the intent distribution points to opportunities for growth in personal and recreational uses of LLMs, which might be overlooked by current research. Observations also indicate these interfaces need to adapt to diverse cultural and linguistic needs.

Note that in the Others option, No one reports valid intents besides our proposed categories. This sideways validates our taxonomy.

\subsubsection{Intents Analysis}

\begin{figure}[htbp]
    \centering
    \includegraphics[width=0.7\linewidth]{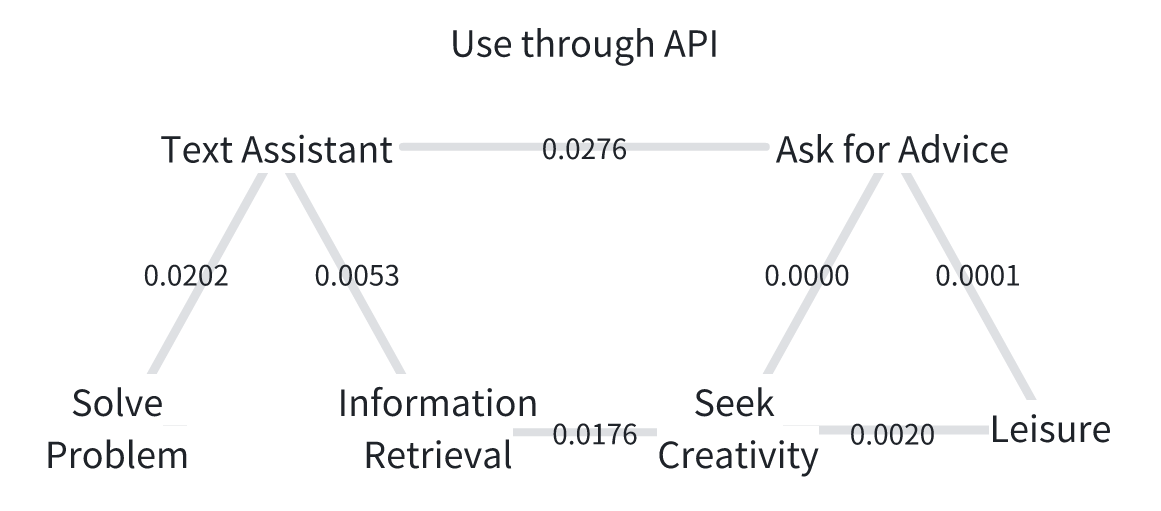}
    % \caption{Independence Examine for Intents. We carried out a chi-square test to analyze whether the users' use of each intention was two-by-two independent and show the pairs with p-value less than 0.05, which indicates the endpoints are not independent. Based on this result, 7 intents are divided into 3 major category: Objective Usage through GUIs, Subjective Usage through GUIs and Usage through APIs.}
    \caption{Pairwise Relationships between Intents: we execute a chi-square test to scrutinize the interdependence of user engagement with each intent. Pairs exhibiting a p-value below 0.05 were identified, signifying a statistically significant correlation. This analytical approach reveals the seven intents distributed across three clusters: Objective Usage through GUIs, Subjective Usage through GUIs, and Usage through APIs.}
    \label{fig:intent-relationship}
\end{figure}

In our study, we explore the interrelationships among user intents based on data indicating whether users engage with each specific intent. To this end, we employ a chi-square test to assess the independence of every pair of intents, analyzing users' engagement patterns across these intent combinations. The outcomes of this examination are depicted in Figure~\ref{fig:intent-relationship}. 

We draw connections, represented by lines, between intent pairs where the chi-square test yields a p-value less than 0.05. This threshold indicates a statistically significant association, suggesting that the connected intents are not independent. 
Then, based on this statistical relevance, we cluster the seven intents into three groups. They are characterized as objective and subjective usage through graphical user interfaces~(GUIs) and usage through application programming interfaces~(APIs). 
% This clustering provides insightful evidence of underlying patterns in how users interact with different aspects of LLMs, revealing a nuanced structure of intent interdependencies.

\medskip
 \fbox{\parbox{0.93\linewidth}{
\textbf{\textit{Finding 2}}: Based on statistical relevance, 7 intents are further clustered into 3 categories: Objective Usage via GUIs, Subjective Usage via GUIs, and Usage through APIs.

\smallskip
\textbf{\textit{Finding 3}}: Text Assistant, Information Retrieval, and Solve Problem in Specialized Areas are the top three usage scenarios.

\smallskip
\textbf{\textit{Finding 4}}: Subjective uses, such as Seeking Creativity and Asking for Advice, are also common intents but may have been overlooked by previous research.
}
}

% \textbf{Findings 4}: \textit{The pattern of Use through API is quite different between Chinese and English participants.}

\subsection{User Satisfaction~(Q5)}
Based on the above analysis of user intents, we further investigate the user self-reported satisfaction in different scenarios and the relationship between satisfaction and usage percentage.
\subsubsection{Rating Analysis}
\label{sec:user-satisfaction}

\begin{figure}[htbp]
    \centering
    \includegraphics[width=0.49\linewidth]{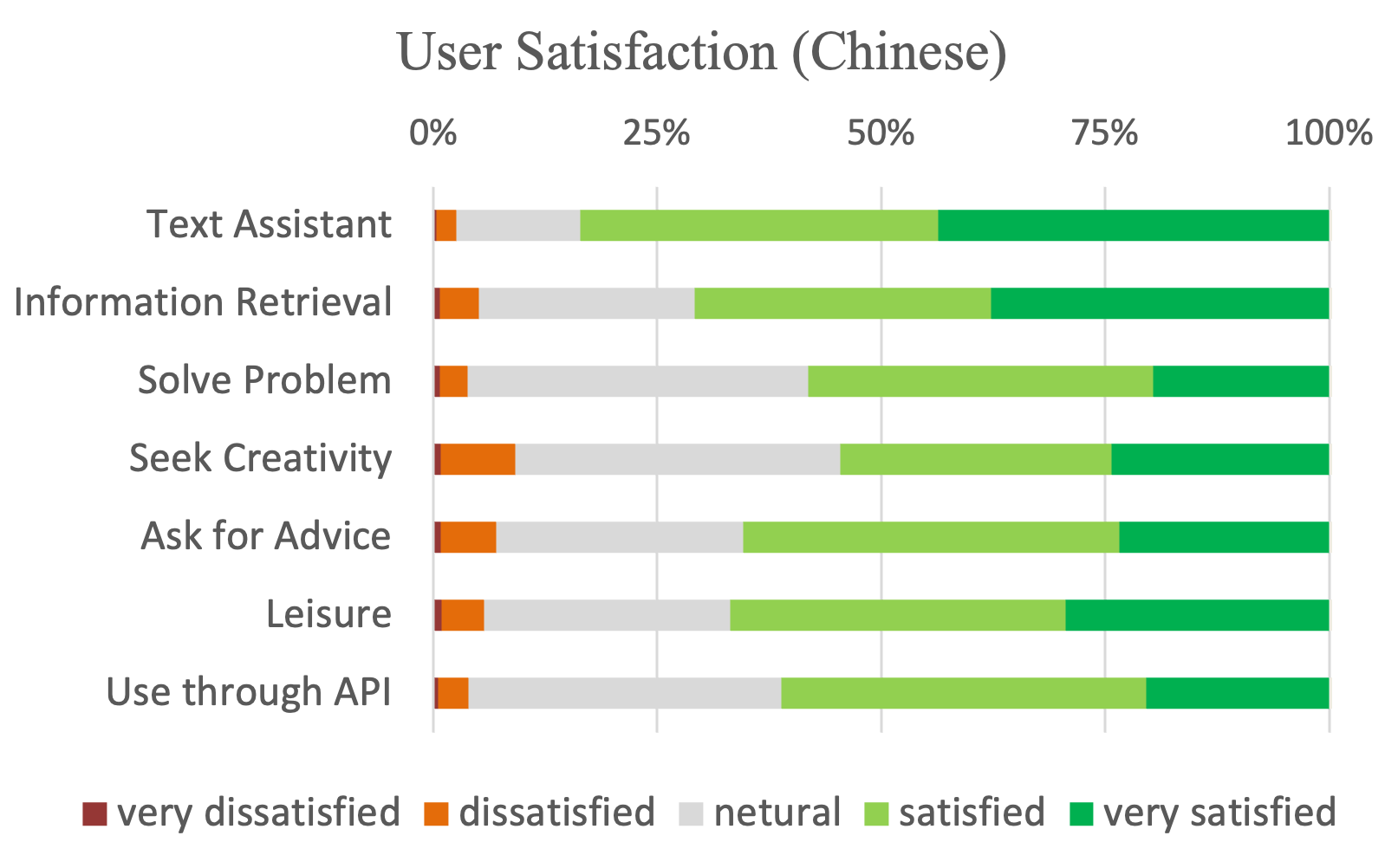}
    \includegraphics[width=0.49\linewidth]{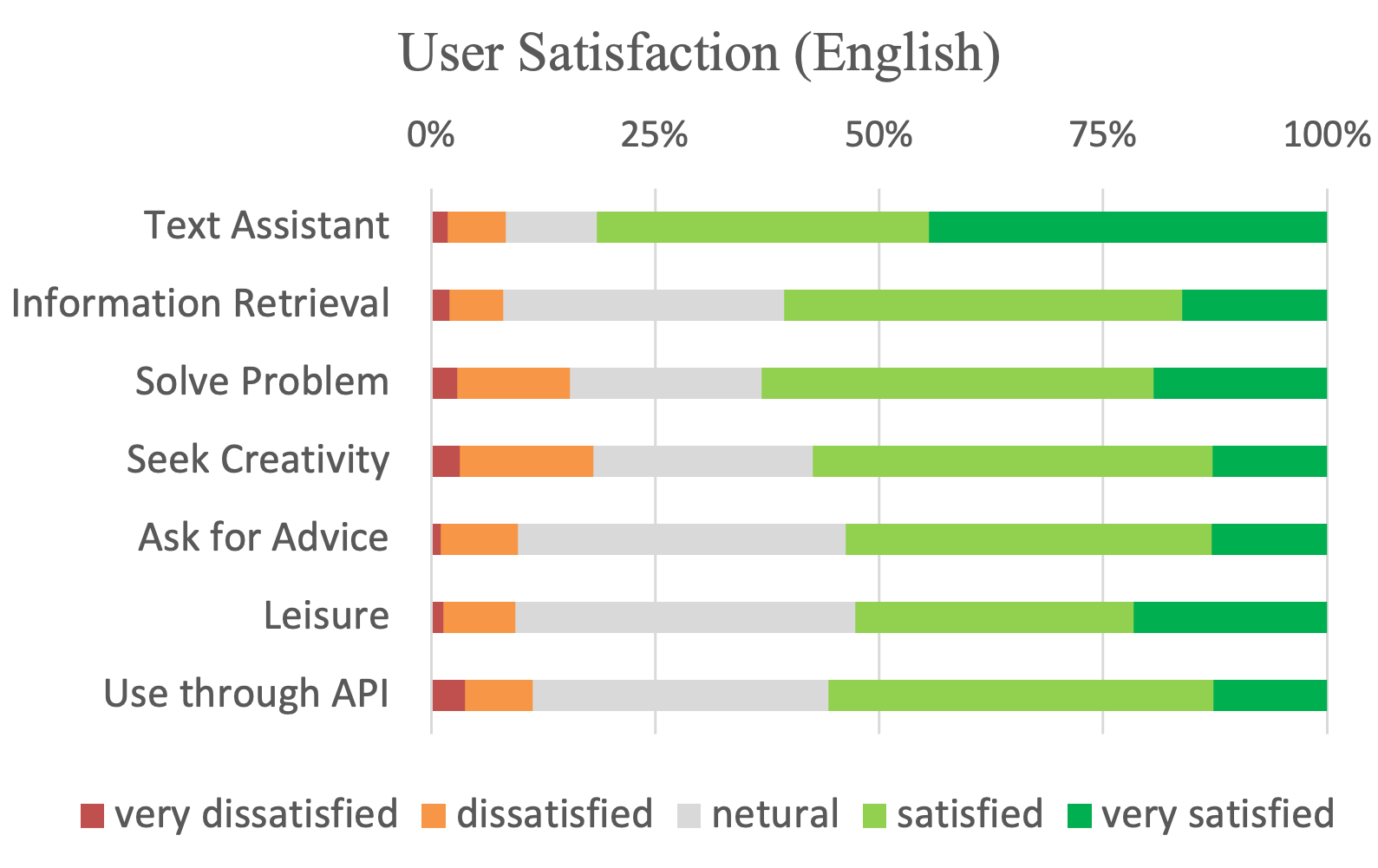}
    \caption{User Satisfaction for engaging with LLMs. We show the ratio of each rating level among users who have used that intent. Vertical coordinates are sorted according to frequency of use in Chinese feedback.}
    \label{fig:user-satisfaction}
\end{figure}

We present the dissatisfaction, neutral, and satisfaction ratios between intents, as shown in Figure~\ref{fig:user-satisfaction}. Note that for each scenario, we only focus on users who have reported using LLMs under this intent, ignoring the N/A feedback.

Here are a few interesting insights from the user satisfaction statistics:

\textbf{Text Assistant} elicited the highest satisfaction rates across both English and Chinese users, with over 80\% reporting being satisfied or very satisfied. This suggests that this conversational service are well-suited to language-based assistance tasks.

\textbf{Seeking Creativity} use cases had the highest negative feedback. Around 9\% of Chinese users and 18\% of English users reported not being satisfied. This indicates that current LLMs have room for improvement when generating novel or imaginative outputs.

While \textbf{Solving Problem} ranked second top for English users, satisfaction was much lower for Chinese at around 58\%. This cross-cultural gap highlights the need for models to be tailored to different contexts to be equally effective globally.

Satisfaction was consistently higher and dissatisfaction lower for most Chinese use cases than in English. This hints at sociocultural norms influencing evaluations to some extent.
% Cultural and Linguistic Differences: 
Besides, the varying degrees of satisfaction between English and Chinese users across different scenarios highlight the impact of cultural and linguistic differences on user experience with LLMs. This suggests a need for more localized and culturally aware services to enhance user satisfaction across the globe.
% More neutral feedback was reported for Problem Solving and Creativity in Chinese feedback compared to other tasks. This could imply these domains involve greater uncertainty that leaves some users unsure how to evaluate experiences.

% In summary, direct exchanges like text manipulating work best, whereas generative and subjective areas require further advances to boost satisfaction across cultures. Accounting for these differences can help optimise LLMs for diverse user experiences.
In summary, textual conversations are effectively supported, but creative and problem-solving capabilities need more tailored development to satisfy diverse global users.
The responses also underscore the importance of considering cultural and linguistic nuances in LLM development and application. Addressing such disparities could further expand the benefit of human-LLM interactions.

% \todo{to be check about this!}

% Conversely, the intents associated with seeking creativity, asking for advice, and leisure activities show a lower satisfaction rate, with a higher proportion of users reporting dissatisfaction. This aligns with the previous analysis, indicating that LLMs may not yet meet the complex needs required for more subjective and creative tasks. The areas with higher reported dissatisfaction could also point to a gap in LLMs' ability to simulate the nuanced human-like interactions that these tasks often require.

% Moreover, the 'N/A' responses were significantly higher for leisure and API use, implying a lack of user experience or a hesitancy to form an opinion on these functions, which could indicate an underdeveloped aspect of LLMs or a lack of user engagement in these areas.

% This satisfaction distribution underscores the need for LLM developers to not only continue enhancing text-related functionalities but also to deepen the models' proficiency in areas requiring personalization, creative cognition, and emotional intelligence to elevate user satisfaction across all intents.

\subsubsection{Satisfaction across Usage Frequency}

% \begin{equation}
% \label{equ:score}
% \text{Rating}(i)=\frac{\sum_{\text{Usage}(u,i)=1}^UR(u,i)}{\sum_{\text{Usage}(u,i)=1}^U1},~
% R(u,i) \in [1,5]
% \end{equation}

\begin{figure}[htbp]
    \centering
    \includegraphics[width=0.49\linewidth]{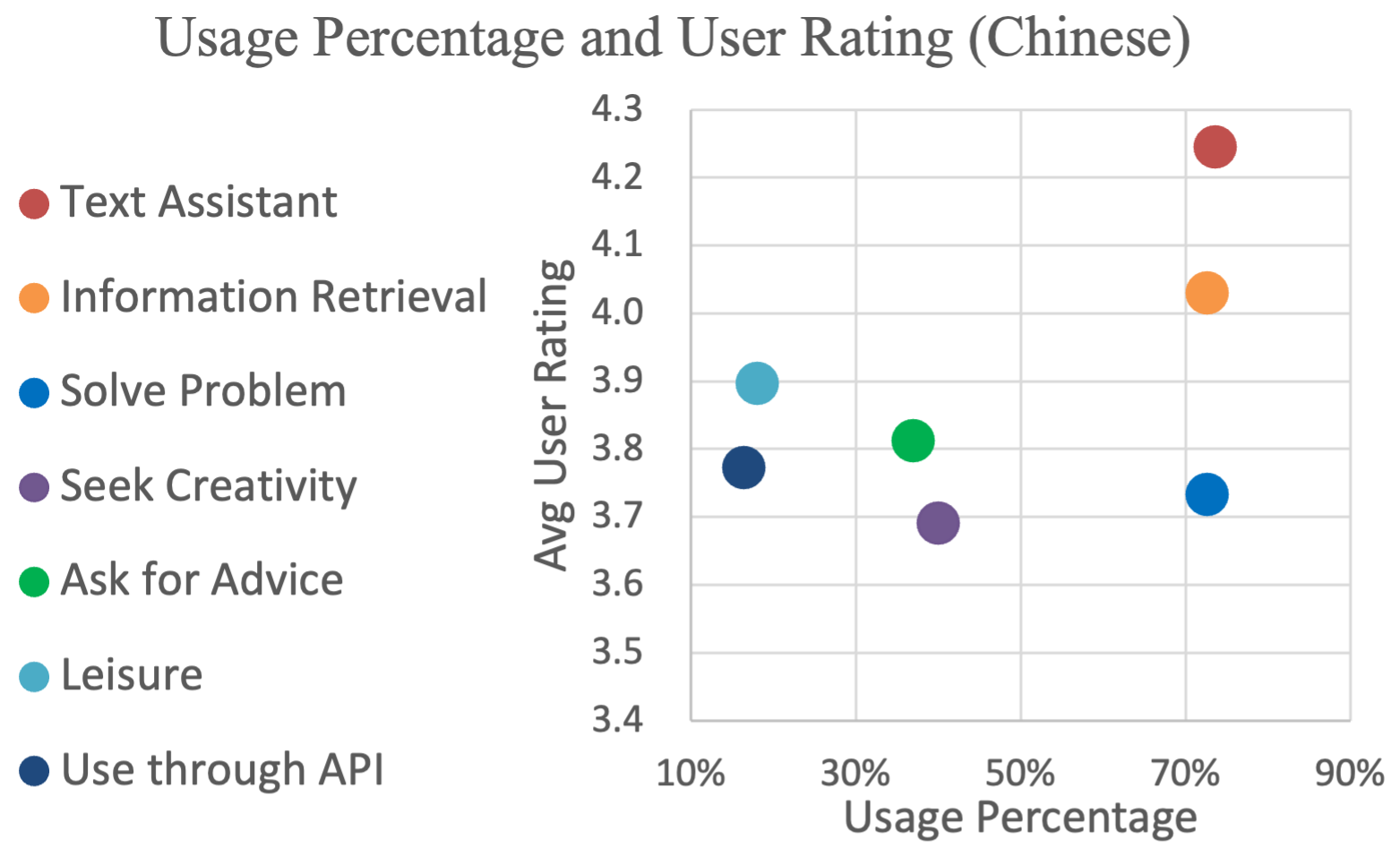}
    \includegraphics[width=0.49\linewidth]{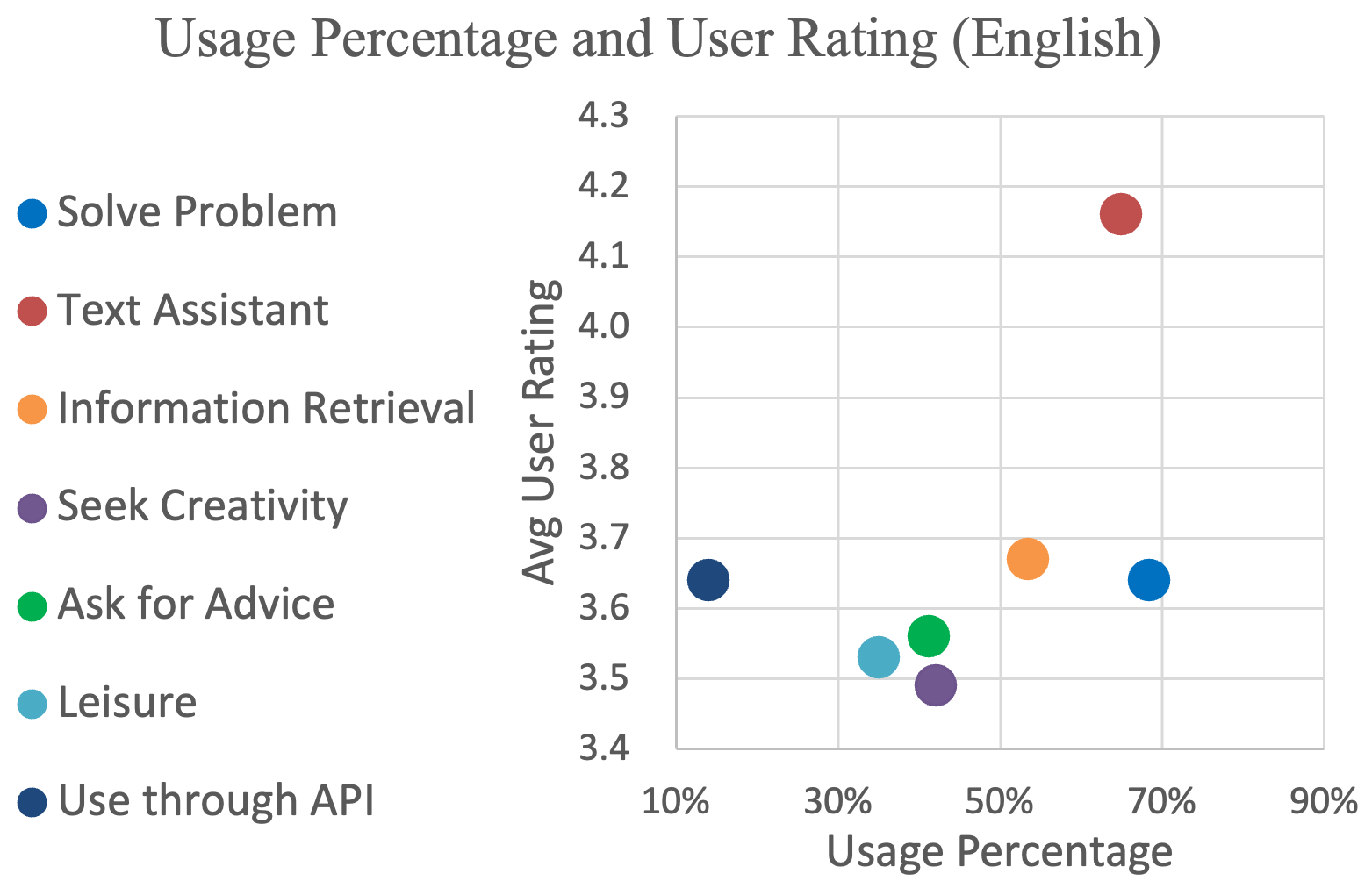}
    \caption{User Usage Percentage and Rating across Intents. Usage percentage represents the ratio of users who used LLMs under the intent.
    User ratings are calculated by assigning 1-5 to "very dissatisfied" - "very satisfied" and then averaging the scores of the users who participated under these intentions. For Chinese feedback,
    when "use through API" is ignored and only use through the GUI is considered, the frequency of use and user satisfaction for each intent is approximately a U-shape.}
    \label{fig:userRating-usage}
\end{figure}

For Chinese feedback, there's a U-shaped correlation between usage frequency and user satisfaction for each intent when ignoring API use and focusing only on GUI interactions. For English feedback, a similar trend of "high on both sides, low in the middle" exists. This implies that users are either satisfied with frequently used features or find less frequently used features surprisingly delightful. Both users and developers need to enhance the visibility and appeal of these hidden gems within LLM platforms, capitalizing on the strengths of lesser-used features and improving medium-frequency collaboration.

\medskip
 \fbox{\parbox{0.93\linewidth}{
\textbf{\textit{Finding 5}}: User studies verify that LLMs are highly effective in text manipulation tasks.

\smallskip
\textbf{\textit{Finding 6}}: Subjective areas, such as Seeking Creativity, require further advances to boost user satisfaction.

\smallskip
\textbf{\textit{Finding 7}}: When both frequencies of use and satisfaction are considered, they approximate a U-shape: both highly and infrequently used scenarios yield higher satisfaction levels.
}
}

\subsection{Expected Response Types~(Q6,7,8)}
\label{sec:response-type}

\begin{figure}[htbp]
    \centering
    \includegraphics[width=0.49\linewidth]{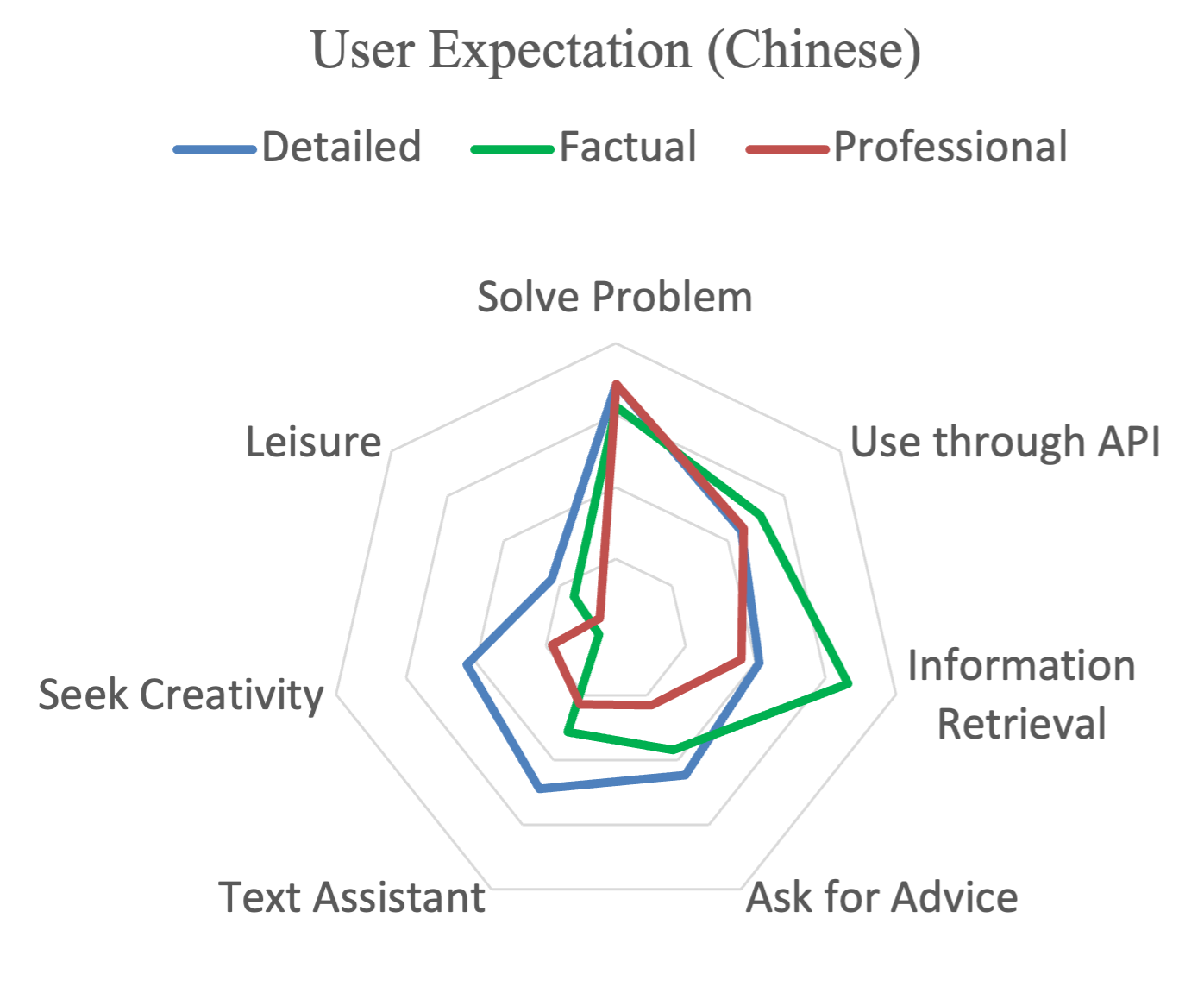}
    \includegraphics[width=0.49\linewidth]{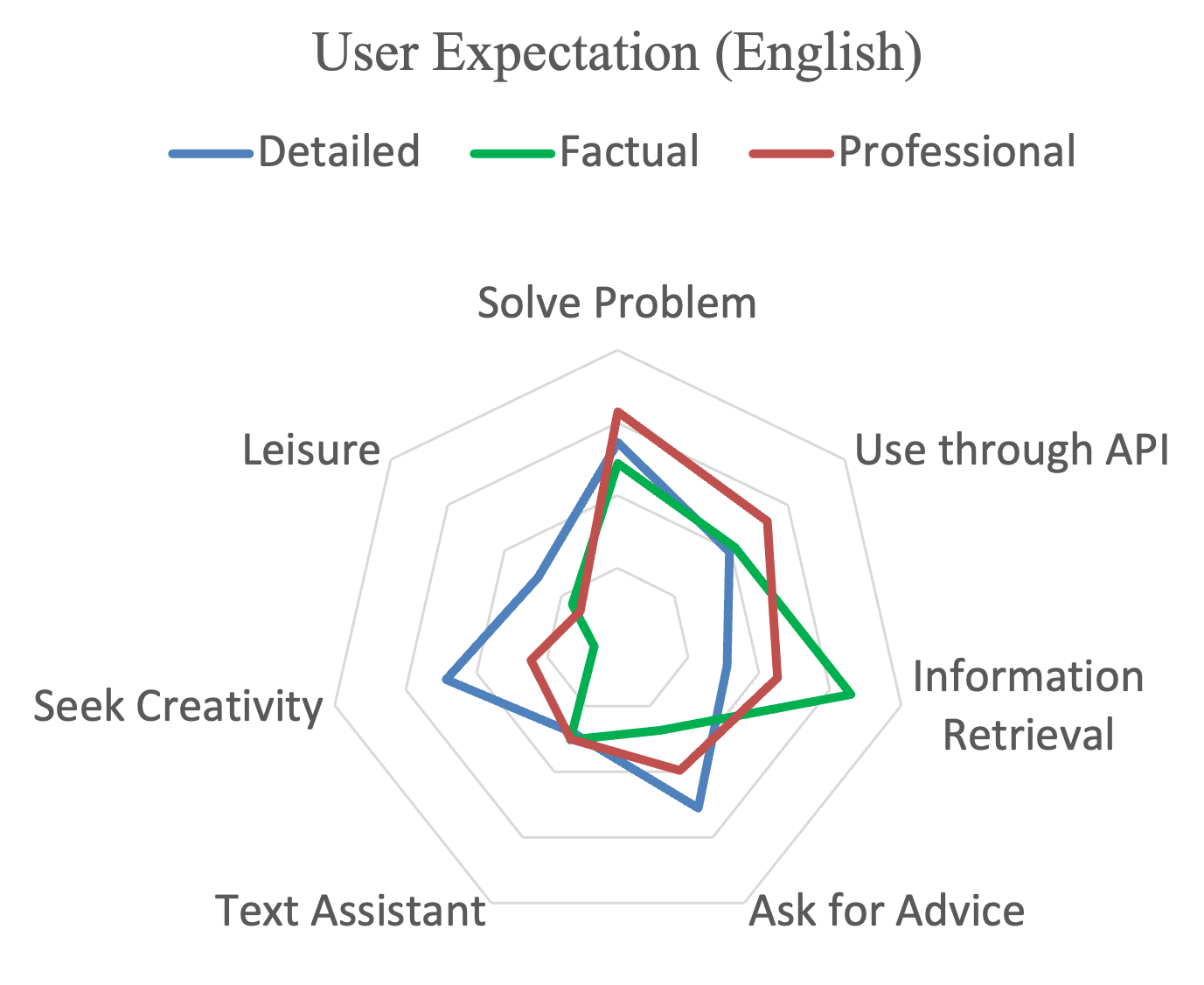}
    \caption{Expected Response Type under Different Intents. Intents are ordered clockwise according to user expectations of the profession. The grid spacing is 25\%, with 0\% at the center and 100\% at the periphery.}
    \label{fig:user-expectation}
\end{figure}

In this section, we analyze user expectations under diverse scenarios.
The results are shown in Figure~\ref{fig:user-expectation}.

Overall, our analysis reveals significant variation in expectations across different intents. Concerning \textbf{response length}, under Leisure scenarios, less than 40\% of both Chinese and English respondents indicate a preference for detailed responses. This trend appears to contradict current reward methodologies that encourage more extended responses~\cite{li2023alpacaeval}. 
This discrepancy highlights a potential misalignment between user preferences and evaluation criteria in certain scenarios.

In the context of \textbf{factual versus creative} responses, less than 40\% of feedback in Ask for Advice, Text Assistant, Leisure, and Seek Creativity favors factual answers. This observation matches the focus on discovering creative capabilities of LLMs~\cite{sinha2023mathematical,shanahan2023evaluating}, as they are predominantly trained to adhere to existing contexts.
When considering \textbf{professional knowledge versus common sense}, only the Solve Problems and Use through API categories exceed a 50\% preference for professional knowledge in both Chinese and English feedback. This trend is partially reflected in current LLM evaluations, which often include professional exams and common sense reasoning benchmarks. However, the former tends to be more emphasized compared to the latter~\cite{achiam2023gpt}, suggesting a possible imbalance with our observed user expectations.

These findings indicate that while LLMs are evaluated on various dimensions, there may be a misalignment with actual user preferences in certain scenarios. Understanding these variances is essential for refining LLMs to better suit real-world user needs in diverse interaction contexts.

% In this section, we discuss user expectations about the interface under different scenarios, ranging from detailed or concise, factual or creative, and with professional knowledge or common sense.
% The results are shown in Figure~\ref{fig:user-expectation}.

% Overall, we can see that the expectations vary greatly across intents.
% From the perspective of context length, both Chinese and English feedback reports lower than 40\% for detailed responses. This may not be in line with the current reward methods that incentivize long responses~\cite{li2023alpacaeval}.
% For factual or creative voting, Ask for Advice, Text Assistant, Leisure, and Seek Creativity all are lower 40\% and there has been related work focus on the creative abilities in LLMs~\cite{sinha2023mathematical,shanahan2023evaluating} as models are trained to follow existing context.
% For Professional Knowledge or Common Sense, only solve problems and Use through API exceed 50\% voting for profession for both Chinese and English. Current evaluations also include professional examinations and common sense reasoning benchmarks~\cite{achiam2023gpt}, while The former tends to have more weight and abundance than the latter, which might align with the observation of user expectations.

% \begin{figure}[htbp]
%     \centering
%     \includegraphics[width=0.33\linewidth]{Figures/Concised-detaled.png}
%       \includegraphics[width=0.33\linewidth]{Figures/Fact-creative.png}
%      \includegraphics[width=0.33\linewidth]{Figures/Professional-common.png}
%     \caption{Expected Response Type (Factual or Creative) under Different Intents. }
%     \label{fig:type-fact}
% \end{figure}

% \begin{figure}[htbp]
%     \centering
%     \includegraphics[width=0.5\linewidth]{Figures/UserExpectation.png}
%     \caption{Overall performance}
%     \label{fig:type-expectation}
% \end{figure}

\smallskip
 \fbox{\parbox{0.93\linewidth}{
\textbf{\textit{Finding 8}}: User Expectations vary greatly across scenarios, which might not always align with the current evaluation standards.
}
}

\subsection{Tool Utilization~(Q9)}
\label{sec:tools}

This section presents the expected tool utilization for users in current LLM services.

\begin{figure}[htbp]
    \centering
    \includegraphics[width=\linewidth]{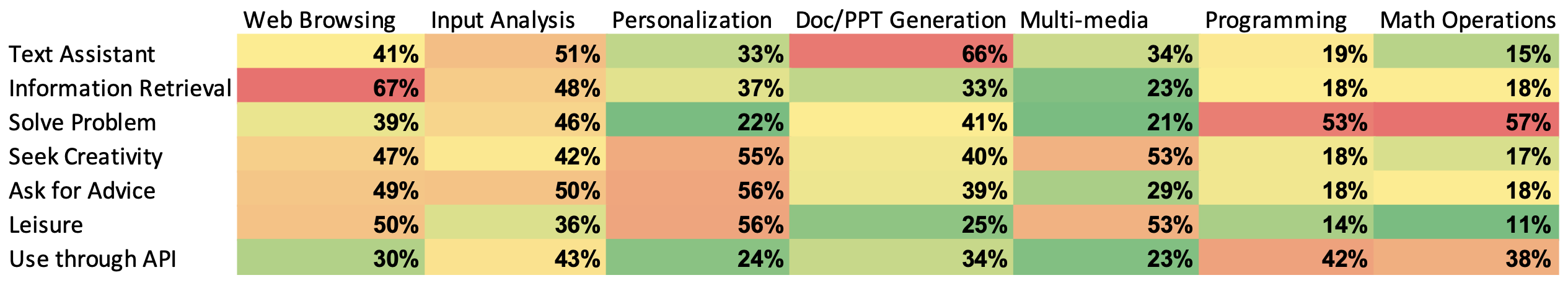}
    \caption{Tool Utilization: users vote for the tools they think are helpful in each scenario. We present the percentage of votes and plot a heat map from red to green based on the magnitude of the values. Since both Chinese and English feedback versions have the same pattern, we show the joint results. Both the horizontal and vertical coordinates are sorted according to overall rating percentages.}
    \label{fig:tool}
\end{figure}

Analyzing the results horizontally, from the intent perspective, reveals a distinct alignment of specific tools with each intent. For instance, \textbf{Text Assistant} commonly utilizes Doc/PPT Generation, \textbf{Information Retrieval} leans on Web Browsing, and \textbf{Problem Solving} frequently employs Programming and Math Operations. This relationship underscores the importance of fine-grained intent categorization, which allows for rapid adaptation of appropriate external tools once the user intent is detected. Such adaptability enhances the relevance of responses and improves the general base model's capability in various scenarios.

Vertically examining the data highlights the significant expectation of Personalization across all subjective scenarios, including \textbf{Seek Creativity}, \textbf{Ask for Advice}, and \textbf{Leisure}. The popularity of recommender systems shows that users expect systems to provide different services even with the same intent as their characteristics and preferences are different. As user-centric services, LLMs need to learn and respond to this dynamic need.

These observations collectively suggest that users expect LLMs to be multifaceted, capable of accurately solving complex professional tasks, and rich in providing personalized or novel responses.

% This data offers valuable insights into user preferences for tool utilization, highlighting opportunities for enhancing LLMs to better cater to diverse user needs in different interaction scenarios.

\medskip
 \fbox{\parbox{0.93\linewidth}{
\textbf{\textit{Finding 9}}: Users anticipate specific tool utilization based on intent, underscoring the necessity of fine-grained scenario segmentation based on user intent.

\smallskip
\textbf{\textit{Finding 10}}: Personalization ability is valued across all subjective usage of LLMs~(Seek Creativity, Ask for Advice, and Leisure).

}
}

\subsection{Major Concerns~(Q11,12)}
\label{sec:conerns}
\begin{figure}[htbp]
    \centering
    \includegraphics[width=0.8\linewidth]{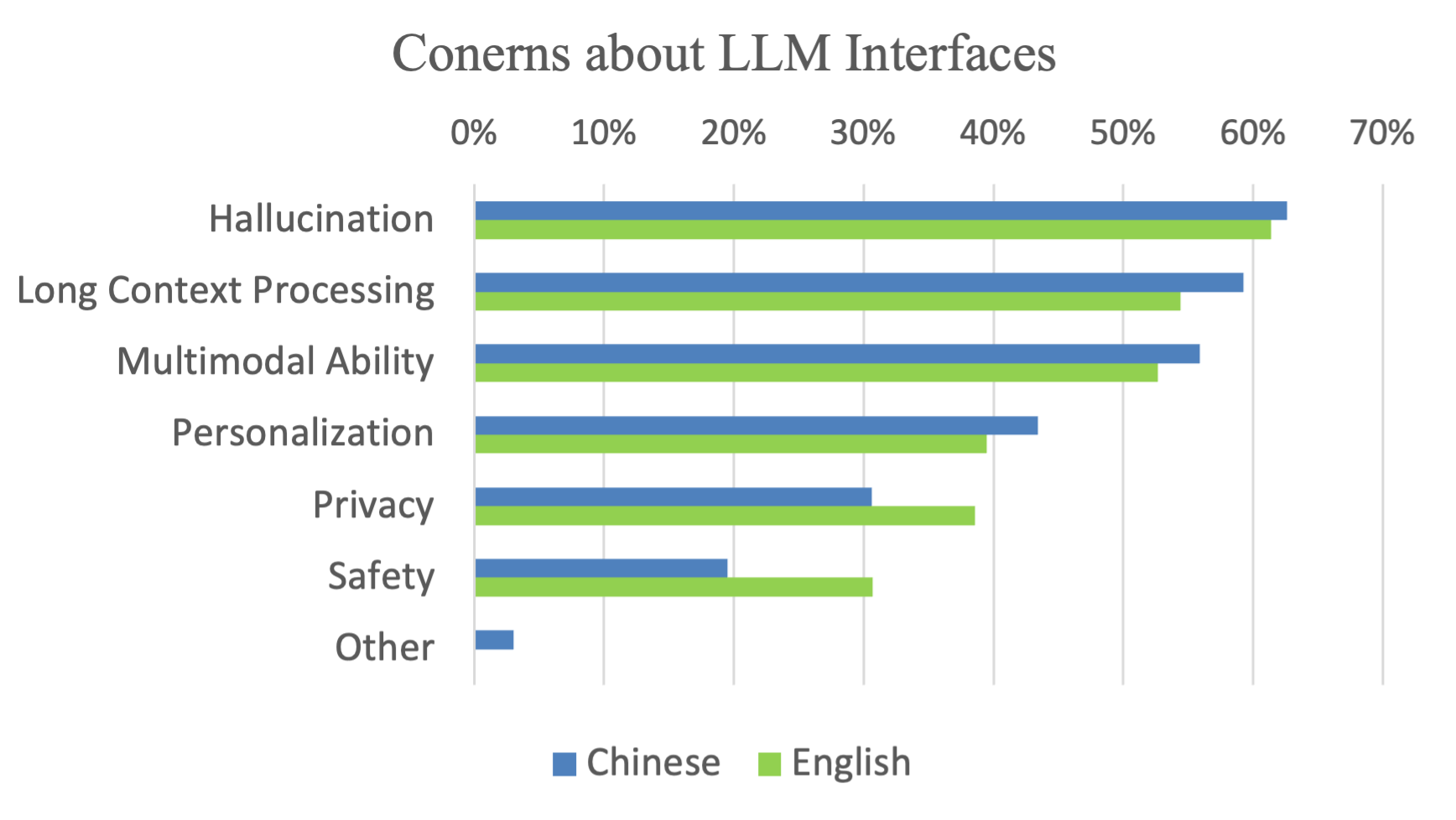}
    \caption{User Concerns and Expected Improvements about current LLM interfaces. Vertical coordinates are sorted according to voting percentages.}
    \label{fig:concerns}
\end{figure}

User concerns and desired improvements for large language models (LLMs) can be broadly classified into two categories: \textbf{Capability} and \textbf{Trustworthiness}. The predominant three concerns, including Hallucination, Long Context Processing, and Multimodal Ability, each surpassing 50\% in user votes, highlight the demand for LLMs to handle diverse and streaming inputs proficiently.

In addition to performance, trustworthiness emerges as another aspect of user concerns. This encompasses ensuring private and safe interactions, which is critical for user confidence and acceptance of this emerging technology. Effectively addressing both these dimensions – capability and trustworthiness – is essential for the successful integration and broader acceptance of LLMs in human-AI collaborations.

% Chinese users expressed the highest concern for hallucination (61.28\%) and long context processing (58.59\%), indicating a strong desire for LLMs to improve in producing factually consistent content and handling lengthy inputs. These concerns suggest that Chinese users place a high value on accuracy and reliability, perhaps reflecting the importance of context continuity and factual integrity in their interactions with LLMs.

% In contrast, English respondents reported a slightly higher rate of concern for hallucination (69.41\%) but less so for long context processing (51.76\%). This may point to a prioritization of content accuracy over the ability to manage long-winded interactions among English-speaking users. Furthermore, the emphasis on privacy (40.00\%) and safety (29.41\%) by English users was notably higher than that of Chinese users, which may reflect cultural or regulatory differences in data protection and content safety standards.

% Multimodal ability concerns were relatively similar for both groups, indicating a shared interest in the LLMs' capabilities to understand and generate multimedia content. The lower concern for personalization among English users compared to Chinese users could suggest differing expectations for tailored content, with Chinese users possibly expecting more adaptive and personalized interactions.

\begin{table}[htbp]
\caption{User self-reported expected improvements in large language models. This is besides the concerns of Hallucination, Long Context Processing, Multimodal Ability, Personalization, Privacy, and Safety.}
\label{table:concerns}
\begin{tabularx}{0.85\textwidth}{ 
  % | >{\raggedright\arraybackslash}X 
  % | >{\centering\arraybackslash}X 
  % | >{\raggedleft\arraybackslash}X 
    >{\hsize=1.5\hsize\linewidth=\hsize}X
    | >{\hsize=.5\hsize\linewidth=\hsize}X
  }
User Concerns in the \textbf{Others} option                                                                                                                              & Type                        \\ \hline
Professionalism, although it can answer my question, but it talks nonsense if I ask for more professional details. So this generic big model is too generic & Professionalism             \\ \hline
Basically unhelpful for specialized fields  & Professionalism   \\ \hline
The dialog is inaccurate, the answers are too generalized, and search engines can find them as well   & Inaccurate,\newline Professionalism \\ \hline
Inaccurate    & Inaccurate   \\ \hline
Logic ability   & Ability (logic)\\ \hline
Does not seem to recognize typos very well in Chinese  & Ability (linguistic)   \\ \hline
Low degree of freedom and many restrictions                                                                                                                  & Freedom                     \\          
\end{tabularx}
\end{table}

Besides the give options illustrated in Figure~\ref{fig:concerns}, there are 8 textual responses under the "Others" option in the Chinese feedback.
After confirming that they were valid and free of sensitive information, the remaining 7 responses are translated into English, shown in table~\ref{table:concerns}. In addition to original user-filled content, we summarize and attach their type.

\medskip
 \fbox{\parbox{0.93\linewidth}{
\textbf{\textit{Finding 11}}: The user concerns and desired improvements are mainly two parts: model capability and trustworthiness.
}
}

\subsection{Summary}
% \todo{unfinished}
In this section, we present the results of the user survey and conclude 11 insightful findings around the current stage of human-LLMs interactions.
In the following section, we proceed to discuss future directions based on this empirical analysis.

% The variance in concerns between the two language groups prompts considerations of cultural, contextual, and perhaps technological influences that shape user expectations and experiences with LLMs. It also invites further investigation into how LLMs can be developed and adapted to meet diverse user needs across different linguistic and cultural landscapes. This understanding is crucial for the global optimization of LLMs, ensuring they are reliable, safe, and sensitive to the nuanced expectations of their varied user base

%% file: section/6_Discussion.tex
\section{Discussion about Future Direction~(RQ 4)}
% Based on the above user studies and empirical analysis, we discuss important issues in future user-centric large language model developments.

\subsection{User-centric Evaluation}
Starting with the evaluation methodologies sets the stage for what follows.
Presently, benchmarks for large language models primarily consist of structured assessments, such as multiple-choice questions or responses evaluated by other reward models~\cite{chang2023survey}. However, it is posited that these test cases and evaluation criteria may not always be congruent with the practical applications and user expectations of generative AI services in real-world usage.

For instance, LLMs are often employed as assistants in both everyday and professional tasks, spanning from objective tasks like information retrieval to more subjective scenarios, including creative or advisory roles, as illustrated in Figure~\ref{fig:intent-relationship}.
While existing benchmarks primarily emphasize subjective usage, focusing predominantly on gauging general intelligence degrees in world knowledge, logical reasoning, and common sense~\cite{achiam2023gpt}, they tend to overlook personal, creative, and leisure-oriented applications. Concerning evaluation criteria, prevailing evaluators such as GPT-4~\cite{li2023alpacaeval} have been noted to exhibit a bias towards longer contexts~\cite{dubois2023alpacafarm}, a tendency that may not consistently align with real-world user preferences, as discussed in Section~\ref{sec:response-type}.

% Thus, advancing evaluations beyond simply ranking LLMs based on their performance in standardized tests or general evaluators, it becomes  to underscore the significance of appraising their applicability and practical utility across a variety of real-life situations.

Therefore, when evaluating generative AI systems, we cannot rely solely on their performance on standardized tests or rankings derived from general evaluators.
Our research posits the necessity for a user-centric evaluation framework. This should encompass test scenarios that are more pertinent to real-world applications and utilize evaluators calibrated to human preferences. Such an approach is imperative for comprehensively appraising of the systems' applicability and practical utility in diverse real-life contexts.

\subsection{User Intent Modeling}
As daily used interfaces, LLMs encounter a variety of usage scenarios.
% , with the expectation of delivering relevant, tailored, and satisfied responses.
Despite the generalist nature of LLMs, the feedback users anticipate can vary markedly, influenced by their specific intent and context. Users may seek different types of responses and tool utilization as delineated in Section~\ref{sec:response-type} and \ref{sec:tools}.
% that are either fact-based or creative, and those that leverage professional expertise or common sense,  among other variations,as delineated in Section~\ref{sec:response-type}.
This expectation of delivering relevant and tailored services raises the research challenge of deciphering different user intents underlying seemingly straightforward inputs to provide further user-centered services.
There are scholarly endeavors focused on intent analysis, which elevate the understanding of user interactions~\cite{shah2022situating,shah2023using}. Moreover, methodologies like Reinforcement Learning with Human Feedback (RLHF)  align generative AI more closely with human preferences~\cite{kaufmann2023survey,lambert2023alignment}. An opportunity arises to refine the adaptability of LLMs further, enabling them to learn from condensed experience and extensive data to recognize and accommodate the diversity inherent in different scenarios.
% Future research needs to augment the adaptability of LLMs. This involves ensuring that these models not only comprehend user intents and expectations from both current queries and historical interactions but also provide pertinent responses tailored to the specific usage scenarios.

% Focusing on LLMs as user-centric services, a key research problem involves understanding the diverse user intents behind simple inputs. As daily-use interfaces, LLMs will encounter a variety of usage scenarios, with the expectation of delivering relevant, tailored, and satisfied responses. 
% % Despite their general-purpose nature, the desired feedback from LLMs greatly varies based on the user's specific intent and the context of the interaction. 
% Although LLMs are general models, the type of feedback users expect can significantly vary depending on their intent and context, such as concise or detailed, based on fact or creation, and use of professional knowledge or common sense, etc, as shown in Section~\ref{sec:response-type}.
% Future research needs to enhance the adaptability of LLMs, ensuring they understand user intents and expectations based on current query and history interactions, and provide relevant responses according to usage scenarios.

\subsection{Personalization}
% Focusing on user intent, the subsequent goal of LLMs is to enable personalization, empowering them to understand and respond to user's personal characteristics.
The widespread adoption of recommendation systems has elucidated that even under identical intents, variances in user preferences can precipitate differences in desired feedback~\cite{lu2015recommender}.
Thus, the subsequent goal of user-centric LLMs is to enable personalization, empowering them to understand and respond better to user's personal characteristics.

% Focusing on user intents, the subsequent objective for LLMs is to advance personalization by enabling them to understand and adapt to individual user characteristics over time. The success of recommendation system have elucidated that enve under identical intents, variances in user preferences can precipitate an expectation of disparate feedback.

The first process of continuous user modeling necessitates the effective handling of streaming inputs~\cite{xiao2023efficient,lin2020limitations,dekel2023speak} and the ability to preserve both immediate and enduring memories or states. Additionally, it calls for advanced methodologies to interpret user preferences and behavioral patterns accurately.
Once user characteristics are understood, the next step is to respond to these insights adaptively.
To reach this degree of adaptability, LLMs need to be structured to dynamically respond to evolving user patterns. Implementing pluggable modules~\cite{zhang2023plug,xiao2023plug} for regular updates and employing efficient tuning mechanisms~\cite{gal2023encoder,paulik2021federated} can facilitate this responsiveness. 

These adjustments are vital for evolving LLMs into dynamic interfaces that comprehend and proactively adapt to user needs, thereby offering a more personalized interaction experience and higher efficiency in human-AI collaborations.

\subsection{Tool Utilization}
Another central research direction is exploring how external tools can be integrated with LLMs to enhance their capabilities and user experience. 
The initial phase entails instructing LLMs with predefined experiences, as illustrated in Section~\ref{sec:tools}, demonstrating the association of specific intents with corresponding tools.
Subsequently, the process evolves to acquiring available tools through self-exploration~\cite{qin2023tool,deitke2022,yang2023holodeck}.
% Then it involves learning to interact freely with the digital and physical world through its own exploratory acquisition~\cite{qin2023tool,deitke2022,yang2023holodeck}.
Moreover, LLMs could progress from the users of external tools to autonomous toolmakers~\cite{cai2023large}, crafting custom utilities to interact freely with the digital and physical world, enabling future embodied intelligence.

\subsection{Trustworthiness}
% Trust and performance are fundamental to the adoption and effective use of LLMs.

Establishing trustworthiness in large language models (LLMs) is paramount, with users expressing concerns about input data privacy and output information reliability, detailed in Section~\ref{sec:conerns}.
To ensure data privacy, strategies like excluding sensitive user data from training~\cite{liu2023trustworthy}, processing messages locally on users' devices~\cite{rahman2023quantized}, or enabling LLMs to unlearn information selectively ~\cite{bourtoule2021machine} are critical. Equally important is data safety, where preventing toxic responses and integrating ethical considerations~\cite{gehman2020realtoxicityprompts} are key measures. Upholding ethical standards in LLM development, focusing on fairness, inclusivity, and bias mitigation, is crucial for gaining user trust and aligning with a user-centric approach.

\subsection{Cross Linguistic and Cultural Development}
The empirical analysis reveals disparities in the user behaviors of English and Chinese-spoken participants, such as usage distributions in information retrieval and APIs, as detailed in Section~\ref{sec:usage-distribution}, and the levels of user satisfaction in asking for advice and leisure, discussed in Section~\ref{sec:user-satisfaction}.
In light of these observations, it becomes incumbent for generative AI to foster both general competence and cross-linguistic and cultural understandings, as discussed in literature~\cite{helm2023diversity,madasu2022large}.

% As shown in the empirical analysis, there have been differences in the usage between English- and Chinese-spoken participants. Such as the usage distribution on Information Retrieval and APIs~in Section\ref{sec:usage-distribution} and the user satisfaction on Ask for Advice and Leisure in Section~\ref{sec:user-satisfaction}.
% It is responsible for generative AI to develop general and cultural abilities~\cite{helm2023diversity,madasu2022large}.

\smallskip
Ultimately, advancing human-LLM collaborations requires a holistic approach beyond mere technical optimizations, including intent understanding and user modeling, self-directed learning of external tools and internal capabilities, enhancing trustworthiness, and user-centered evaluations for real-world utilities, etc.
Tackling these multifaceted challenges is essential for building effective and responsible LLMs to propel human-AI collaborations.

% Establishing trustworthiness in LLMs remains a critical challenge. Users often have concerns about the privacy of input information and the reliability of output data.

% Addressing data privacy involves strategies such as excluding sensitive user data from model training, localizing some processing to the user side, or enabling LLMs to unlearn specific information. For data safety, it's essential to prevent toxic responses and incorporate further ethical considerations.
% Commitments to ethical aspects in LLM development, such as fairness, inclusivity, and bias mitigation, are essential for building trust and fostering user acceptance, aligning with a core user-centric approach.

% In conclusion, the progression in user-LLM interaction demands a comprehensive approach that goes beyond technical prowess. It encompasses facets of enhancing its core abilities, better user-centric understanding and trustworthiness. Addressing these diverse aspects is key to driving future research and achieving more responsive, effective, and ethically responsible user-centric AI systems.

% \subsection{Cross Linguistic and Cultural Development}

%% file: section/7.tex
\section{Limitations}
\label{sec:limitation}

The participant demographics of our user survey, primarily disseminated through social media of graduate students and university professors, suggest a likely skew towards highly educated individuals. This user group, potentially more experienced in human-AI collaboration, may have a predisposition to engage with and deeply utilize new technologies. Additionally, the voluntary nature of the survey participation introduces a selection bias, as it tends to attract users with a pre-existing interest in LLMs or those inclined to contribute to anonymous user studies.

Given these factors, our study is not representative of a random, general user group but rather a more specific segment that potentially has deep and sophisticated engagements with LLMs.
% , potentially leading to high quality feedback from users with deep and sophisticated engagements with LLM interfaces. 
While this offers valuable insights, it's important to acknowledge this limitation regarding the broader applicability and generalization of the findings.

% Chinese VS English.
% Spread through social networks and the Internet for mainly university students and professors.
% Some results remain steady across Chinese and English versions. that reflects some steady information.

% \todo{selection bias}
% users who do not use large language models might not pay attention to the user survey compared to users who engage with LLMs a lot.

\section{Conclusion}
% In this study, we argue for a user-centric perspective in large language model evaluation and development. By doing so, we first extend the user intent studies in information-seeking services to the scope of general large language model-driven interfaces, defining a taxonomy with 7 intents that fail in three major categories: subjective use through GUI interfaces, objective use through GUI interfaces, and usage through API. 
% With this intent understanding, we can conduct fine-grained user-centric evaluation of LLMs that align closely with their utility in real-world human-AI collaborations.
% Then, we conduct a user study and collect 411 anonymous feedback about the first-hand usage frequency, user experience, and concerns about current LLMs. This data will be publicly released alongside the publication of this paper.
% We draw \todo{xx} key findings from this survey, including validation of LLMs are widely adopted on a daily basis and some use cases might be overlooked by current studies.
% With these insights from empirical analysis, we discuss and highlight \todo{xxx} future direction with advancing LLMs for better human-AI collaboration instead of mere technique improvements.

% 2 fine-grained, use user intents to divide

% 3 Among the intents, Subjective scenario like Seeking Creativity, Asking for advice and Leisure, are most ignoring compared to the objective usage.

In this study, we embrace a user-centric approach for evaluating large language models (LLMs). 
% This approach initially directs us to a profound comprehension of user intents in the realm of LLM interactions.
This perspective steers us toward an in-depth understanding of user intents.
We broaden the scope of existing classifications from traditional information-seeking systems to general LLM-driven interfaces, formulating a taxonomy of 7 intents.
This refined classification, grounded by open-source ChatGPT user logs and supplemented with human verification, sets the stage for fine-grained explorations of user satisfaction with LLMs.

Building on this understanding,
we conducted a user study, collecting anonymous feedback from 411 individuals. This feedback spans aspects of usage frequency, experience corresponding to each intent, and concerns regarding LLMs. This insightful dataset will be released upon the publication of our paper.
Analysis of this survey yielded 11 compelling findings, starting with the confirmation of the widespread daily use of the LLM.
The proposed intents are categorized into three statistically relevant clusters: subjective uses through graphical user interfaces~(GUIs), objective uses through GUIs, and usage through application programming interfaces~(APIs). 
Notably, subjective usage through GUIs are potentially overlooked by existing research. 
The observations derived from empirical analysis further guide us in discussions of future directions for advancing LLMs, focusing not just on technical optimizations but on augmenting human-AI collaboration. This shift towards a more user-centric development of LLMs is crucial for ensuring that these models meet the diverse and evolving needs in real-world interactions.

\section*{Acknowledgement}
Thanks to all the participants for taking the user study and giving valuable feedback about the first-hand experience with the large language model interfaces.

%% file: appendix.tex
\section*{Appendix}
\section{Questionnaire}
\label{sec:appendix}

\begin{enumerate}[leftmargin=*,nosep]
    \item Please select the services you have used (in Descending Order of Frequency) [Order]\\
    Options are sorted alphabetically
    \item Your frequency of usage [Single Choice]\\
    Options: Daily, Weekly, Monthly, Tried, Never.\\

    \item What are the following do you use the system for? [Multi Choices]\\ 
    Detailed explanations about these categories are included, which is the same as Section~\ref{subsec:intents}.\\
    Ask for Advice,
    Information Retrieval, 
    Leisure,
    Seek Creativity,  
    Solve Problems in Specialized Areas, 
    Text Assistant, 
    Use through API,
    Others~[fill in].
    \item Do you have anything to add or comment to the categories above? [Blank]\\

    \item For the following usage scenarios, your rating on the system performance is? [Single Choice]\\
    If you have not used the system for certain requirements, please select N/A.\\
    For each intent, the options are: Very dissatisfied, dissatisfied, neutral, satisfied, very satisfied, N/A.\\
    
    \item For the following usage scenarios, do you think the system's response needs to be [concise] or [detailed]? [Single Choice * 7 intents]
    \item For the following usage scenarios, do you think the system's answers need to be [based on facts] or [based on creations]? [Single Choice * 7 intents]
    \item For the following usage scenarios, do you think the system's answers need to be [based on professional knowledge] or [based on common sense]? [Single Choice * 7 intents]\\
    
    \item In each scenario, which of the following tools helps the system generate higher-quality responses? [Multi Choices * 7 intents]\\
    Web Browsing: use search engines, browse and analyze web content, etc.\\
    Input Analysis: understand uploaded documents, pictures, etc.\\
    Personalization: recommend content based on user interests\\
    Programming: write and execute code\\
    Mathematical Operations: Assist with mathematical calculations, logical reasoning \\
    Documentation Generation: create diagrams, documents and slides \\
    Multimedia Creation: generate pictures and videos\\
    Other tools: please input\\
    
    % \item Please briefly describe 3-5 recent conversations you have had with the Generative Al system [Choice and blank]\\
    % You could refer to the title generated by the system, or summarize by yourself\\
    % Please fill in the description in the corresponding category, if you cannot determine please fill in the others\\
    \item Please select B for this question, thanks! [Single Choice]\\

    \item Which aspects of the existing system do you think are lacking and most in need of optimization? [Multi Choice]\\
    Hallucination, Long context processing, multimodal understanding and generation, personalization, privacy, safety, others~[fill in].
    \item Do you have other comments and suggestions about this questionnaire or large language model services? Thanks for your participation and support! [Blank]

\end{enumerate}

\clearpage
\section{Demographic Information}
\label{sec:IPs}

\begin{figure}[htbp]
    \centering
    \includegraphics[width=0.7\linewidth]{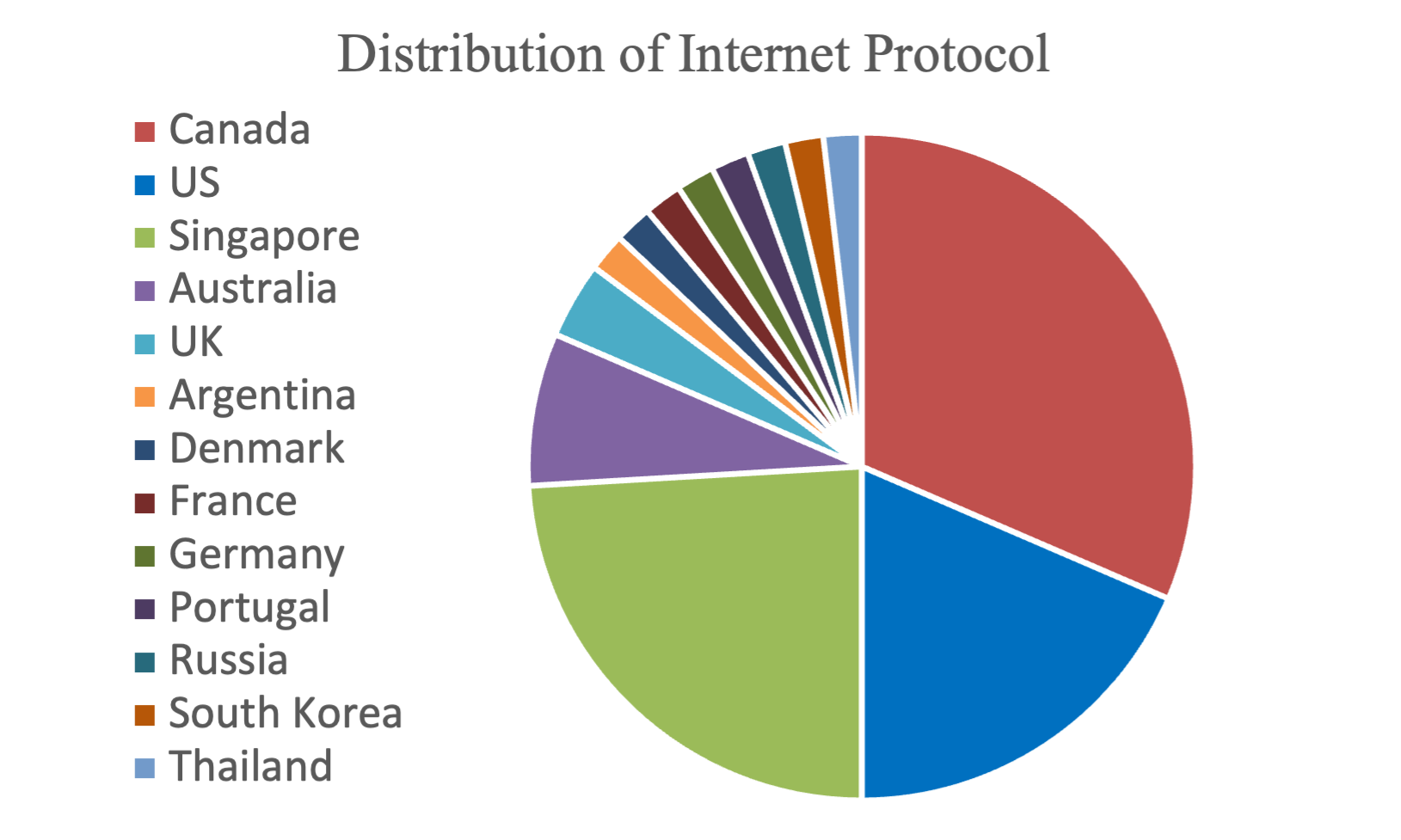}
    \caption{IP distribution excluding China.}
    \label{fig:IPs}
\end{figure}

In our survey, personal questions were not included. The sole demographic information obtained was the automatically recorded IP addresses captured by the questionnaire system. 
For the Chinese version of survey, we recorded 291 IPs originating from China and 6 from other countries. In the English version of the survey, there were 66 Chinese IPs and 48 IPs from various other countries. It is pertinent to note that the English survey was disseminated amongst an international graduate program based in China, resulting in approximately 30 responses from English-speaking students located within China. The distribution of IP addresses from outside China is depicted in Figure~\ref{fig:IPs}.

%% file: main.bbl

\begin{thebibliography}{73}

%%% the Web address.

\ifx \showCODEN    \undefined \def \showCODEN     #1{\unskip}     \fi
\ifx \showDOI      \undefined \def \showDOI       #1{#1}\fi
\ifx \showISBNx    \undefined \def \showISBNx     #1{\unskip}     \fi
\ifx \showISBNxiii \undefined \def \showISBNxiii  #1{\unskip}     \fi
\ifx \showISSN     \undefined \def \showISSN      #1{\unskip}     \fi
\ifx \showLCCN     \undefined \def \showLCCN      #1{\unskip}     \fi
\ifx \shownote     \undefined \def \shownote      #1{#1}          \fi
\ifx \showarticletitle \undefined \def \showarticletitle #1{#1}   \fi
\ifx \showURL      \undefined \def \showURL       {\relax}        \fi
% invisible to TeX
\providecommand\bibfield[2]{#2}
\providecommand\bibinfo[2]{#2}
\providecommand\natexlab[1]{#1}
\providecommand\showeprint[2][]{arXiv:#2}

\bibitem[Achiam et~al\mbox{.}(2023)]%
        {achiam2023gpt}
\bibfield{author}{\bibinfo{person}{Josh Achiam}, \bibinfo{person}{Steven Adler}, \bibinfo{person}{Sandhini Agarwal}, \bibinfo{person}{Lama Ahmad}, \bibinfo{person}{Ilge Akkaya}, \bibinfo{person}{Florencia~Leoni Aleman}, \bibinfo{person}{Diogo Almeida}, \bibinfo{person}{Janko Altenschmidt}, \bibinfo{person}{Sam Altman}, \bibinfo{person}{Shyamal Anadkat}, {et~al\mbox{.}}} \bibinfo{year}{2023}\natexlab{}.
\newblock \showarticletitle{GPT-4 Technical Report}.
\newblock \bibinfo{journal}{\emph{arXiv preprint}} (\bibinfo{year}{2023}).
\newblock


\bibitem[Bang et~al\mbox{.}(2023)]%
        {bang2023multitask}
\bibfield{author}{\bibinfo{person}{Yejin Bang}, \bibinfo{person}{Samuel Cahyawijaya}, \bibinfo{person}{Nayeon Lee}, \bibinfo{person}{Wenliang Dai}, \bibinfo{person}{Dan Su}, \bibinfo{person}{Bryan Wilie}, \bibinfo{person}{Holy Lovenia}, \bibinfo{person}{Ziwei Ji}, \bibinfo{person}{Tiezheng Yu}, \bibinfo{person}{Willy Chung}, {et~al\mbox{.}}} \bibinfo{year}{2023}\natexlab{}.
\newblock \showarticletitle{A multitask, multilingual, multimodal evaluation of chatgpt on reasoning, hallucination, and interactivity}.
\newblock \bibinfo{journal}{\emph{arXiv preprint arXiv:2302.04023}} (\bibinfo{year}{2023}).
\newblock


\bibitem[Bendersky and Croft(2009)]%
        {bendersky2009analysis}
\bibfield{author}{\bibinfo{person}{Michael Bendersky} {and} \bibinfo{person}{W~Bruce Croft}.} \bibinfo{year}{2009}\natexlab{}.
\newblock \showarticletitle{Analysis of long queries in a large scale search log}. In \bibinfo{booktitle}{\emph{Proceedings of the 2009 workshop on Web Search Click Data}}. \bibinfo{pages}{8--14}.
\newblock


\bibitem[Bodroza et~al\mbox{.}(2023)]%
        {bodroza2023personality}
\bibfield{author}{\bibinfo{person}{Bojana Bodroza}, \bibinfo{person}{Bojana~M Dinic}, {and} \bibinfo{person}{Ljubisa Bojic}.} \bibinfo{year}{2023}\natexlab{}.
\newblock \showarticletitle{Personality testing of GPT-3: Limited temporal reliability, but highlighted social desirability of GPT-3's personality instruments results}.
\newblock \bibinfo{journal}{\emph{arXiv preprint arXiv:2306.04308}} (\bibinfo{year}{2023}).
\newblock


\bibitem[Bourtoule et~al\mbox{.}(2021)]%
        {bourtoule2021machine}
\bibfield{author}{\bibinfo{person}{Lucas Bourtoule}, \bibinfo{person}{Varun Chandrasekaran}, \bibinfo{person}{Christopher~A Choquette-Choo}, \bibinfo{person}{Hengrui Jia}, \bibinfo{person}{Adelin Travers}, \bibinfo{person}{Baiwu Zhang}, \bibinfo{person}{David Lie}, {and} \bibinfo{person}{Nicolas Papernot}.} \bibinfo{year}{2021}\natexlab{}.
\newblock \showarticletitle{Machine unlearning}. In \bibinfo{booktitle}{\emph{2021 IEEE Symposium on Security and Privacy (SP)}}. IEEE, \bibinfo{pages}{141--159}.
\newblock


\bibitem[Broder(2002)]%
        {broder2002taxonomy}
\bibfield{author}{\bibinfo{person}{Andrei Broder}.} \bibinfo{year}{2002}\natexlab{}.
\newblock \showarticletitle{A taxonomy of web search}. In \bibinfo{booktitle}{\emph{ACM Sigir forum}}, Vol.~\bibinfo{volume}{36}. ACM New York, NY, USA, \bibinfo{pages}{3--10}.
\newblock


\bibitem[Bubeck et~al\mbox{.}(2023)]%
        {bubeck2023sparks}
\bibfield{author}{\bibinfo{person}{S{\'e}bastien Bubeck}, \bibinfo{person}{Varun Chandrasekaran}, \bibinfo{person}{Ronen Eldan}, \bibinfo{person}{Johannes Gehrke}, \bibinfo{person}{Eric Horvitz}, \bibinfo{person}{Ece Kamar}, \bibinfo{person}{Peter Lee}, \bibinfo{person}{Yin~Tat Lee}, \bibinfo{person}{Yuanzhi Li}, \bibinfo{person}{Scott Lundberg}, {et~al\mbox{.}}} \bibinfo{year}{2023}\natexlab{}.
\newblock \showarticletitle{Sparks of artificial general intelligence: Early experiments with gpt-4}.
\newblock \bibinfo{journal}{\emph{arXiv preprint arXiv:2303.12712}} (\bibinfo{year}{2023}).
\newblock


\bibitem[Cai et~al\mbox{.}(2023)]%
        {cai2023large}
\bibfield{author}{\bibinfo{person}{Tianle Cai}, \bibinfo{person}{Xuezhi Wang}, \bibinfo{person}{Tengyu Ma}, \bibinfo{person}{Xinyun Chen}, {and} \bibinfo{person}{Denny Zhou}.} \bibinfo{year}{2023}\natexlab{}.
\newblock \showarticletitle{Large language models as tool makers}.
\newblock \bibinfo{journal}{\emph{arXiv preprint arXiv:2305.17126}} (\bibinfo{year}{2023}).
\newblock


\bibitem[Chang et~al\mbox{.}(2023)]%
        {chang2023survey}
\bibfield{author}{\bibinfo{person}{Yupeng Chang}, \bibinfo{person}{Xu Wang}, \bibinfo{person}{Jindong Wang}, \bibinfo{person}{Yuan Wu}, \bibinfo{person}{Kaijie Zhu}, \bibinfo{person}{Hao Chen}, \bibinfo{person}{Linyi Yang}, \bibinfo{person}{Xiaoyuan Yi}, \bibinfo{person}{Cunxiang Wang}, \bibinfo{person}{Yidong Wang}, {et~al\mbox{.}}} \bibinfo{year}{2023}\natexlab{}.
\newblock \showarticletitle{A survey on evaluation of large language models}.
\newblock \bibinfo{journal}{\emph{arXiv preprint arXiv:2307.03109}} (\bibinfo{year}{2023}).
\newblock


\bibitem[Chen et~al\mbox{.}(2012)]%
        {chen2012understanding}
\bibfield{author}{\bibinfo{person}{Long Chen}, \bibinfo{person}{Dell Zhang}, {and} \bibinfo{person}{Levene Mark}.} \bibinfo{year}{2012}\natexlab{}.
\newblock \showarticletitle{Understanding user intent in community question answering}. In \bibinfo{booktitle}{\emph{Proceedings of the 21st international conference on world wide web}}. \bibinfo{pages}{823--828}.
\newblock


\bibitem[Chen et~al\mbox{.}(2021)]%
        {chen2021evaluating}
\bibfield{author}{\bibinfo{person}{Mark Chen}, \bibinfo{person}{Jerry Tworek}, \bibinfo{person}{Heewoo Jun}, \bibinfo{person}{Qiming Yuan}, \bibinfo{person}{Henrique Ponde de~Oliveira Pinto}, \bibinfo{person}{Jared Kaplan}, \bibinfo{person}{Harri Edwards}, \bibinfo{person}{Yuri Burda}, \bibinfo{person}{Nicholas Joseph}, \bibinfo{person}{Greg Brockman}, {et~al\mbox{.}}} \bibinfo{year}{2021}\natexlab{}.
\newblock \showarticletitle{Evaluating large language models trained on code}.
\newblock \bibinfo{journal}{\emph{arXiv preprint arXiv:2107.03374}} (\bibinfo{year}{2021}).
\newblock


\bibitem[Choi and Schwarcz(2023)]%
        {choi2023ai}
\bibfield{author}{\bibinfo{person}{Jonathan~H Choi} {and} \bibinfo{person}{Daniel Schwarcz}.} \bibinfo{year}{2023}\natexlab{}.
\newblock \showarticletitle{AI assistance in legal analysis: An empirical study}.
\newblock  (\bibinfo{year}{2023}).
\newblock


\bibitem[Chu et~al\mbox{.}(2022)]%
        {chu2022empirical}
\bibfield{author}{\bibinfo{person}{Hyeshin Chu}, \bibinfo{person}{Joohee Kim}, \bibinfo{person}{Seongouk Kim}, \bibinfo{person}{Hongkyu Lim}, \bibinfo{person}{Hyunwook Lee}, \bibinfo{person}{Seungmin Jin}, \bibinfo{person}{Jongeun Lee}, \bibinfo{person}{Taehwan Kim}, {and} \bibinfo{person}{Sungahn Ko}.} \bibinfo{year}{2022}\natexlab{}.
\newblock \showarticletitle{An Empirical Study on How People Perceive AI-generated Music}. In \bibinfo{booktitle}{\emph{Proceedings of the 31st ACM International Conference on Information \& Knowledge Management}}.
\newblock


\bibitem[Clark et~al\mbox{.}(2018)]%
        {clark2018think}
\bibfield{author}{\bibinfo{person}{Peter Clark}, \bibinfo{person}{Isaac Cowhey}, \bibinfo{person}{Oren Etzioni}, \bibinfo{person}{Tushar Khot}, \bibinfo{person}{Ashish Sabharwal}, \bibinfo{person}{Carissa Schoenick}, {and} \bibinfo{person}{Oyvind Tafjord}.} \bibinfo{year}{2018}\natexlab{}.
\newblock \showarticletitle{Think you have solved question answering? try arc, the ai2 reasoning challenge}.
\newblock \bibinfo{journal}{\emph{arXiv preprint}} (\bibinfo{year}{2018}).
\newblock


\bibitem[Cobbe et~al\mbox{.}(2021)]%
        {cobbe2021training}
\bibfield{author}{\bibinfo{person}{Karl Cobbe}, \bibinfo{person}{Vineet Kosaraju}, \bibinfo{person}{Mohammad Bavarian}, \bibinfo{person}{Mark Chen}, \bibinfo{person}{Heewoo Jun}, \bibinfo{person}{Lukasz Kaiser}, \bibinfo{person}{Matthias Plappert}, \bibinfo{person}{Jerry Tworek}, \bibinfo{person}{Jacob Hilton}, \bibinfo{person}{Reiichiro Nakano}, {et~al\mbox{.}}} \bibinfo{year}{2021}\natexlab{}.
\newblock \showarticletitle{Training verifiers to solve math word problems}.
\newblock \bibinfo{journal}{\emph{arXiv preprint arXiv:2110.14168}} (\bibinfo{year}{2021}).
\newblock


\bibitem[Dai et~al\mbox{.}(2023b)]%
        {dai2023uncovering}
\bibfield{author}{\bibinfo{person}{Sunhao Dai}, \bibinfo{person}{Ninglu Shao}, \bibinfo{person}{Haiyuan Zhao}, \bibinfo{person}{Weijie Yu}, \bibinfo{person}{Zihua Si}, \bibinfo{person}{Chen Xu}, \bibinfo{person}{Zhongxiang Sun}, \bibinfo{person}{Xiao Zhang}, {and} \bibinfo{person}{Jun Xu}.} \bibinfo{year}{2023}\natexlab{b}.
\newblock \showarticletitle{Uncovering ChatGPT's Capabilities in Recommender Systems}.
\newblock \bibinfo{journal}{\emph{arXiv preprint arXiv:2305.02182}} (\bibinfo{year}{2023}).
\newblock


\bibitem[Dai et~al\mbox{.}(2023a)]%
        {dai2023can}
\bibfield{author}{\bibinfo{person}{Wei Dai}, \bibinfo{person}{Jionghao Lin}, \bibinfo{person}{Hua Jin}, \bibinfo{person}{Tongguang Li}, \bibinfo{person}{Yi-Shan Tsai}, \bibinfo{person}{Dragan Ga{\v{s}}evi{\'c}}, {and} \bibinfo{person}{Guanliang Chen}.} \bibinfo{year}{2023}\natexlab{a}.
\newblock \showarticletitle{Can large language models provide feedback to students? A case study on ChatGPT}. In \bibinfo{booktitle}{\emph{2023 IEEE International Conference on Advanced Learning Technologies (ICALT)}}. IEEE, \bibinfo{pages}{323--325}.
\newblock


\bibitem[Deitke et~al\mbox{.}(2022)]%
        {deitke2022}
\bibfield{author}{\bibinfo{person}{Matt Deitke}, \bibinfo{person}{Eli VanderBilt}, \bibinfo{person}{Alvaro Herrasti}, \bibinfo{person}{Luca Weihs}, \bibinfo{person}{Kiana Ehsani}, \bibinfo{person}{Jordi Salvador}, \bibinfo{person}{Winson Han}, \bibinfo{person}{Eric Kolve}, \bibinfo{person}{Aniruddha Kembhavi}, {and} \bibinfo{person}{Roozbeh Mottaghi}.} \bibinfo{year}{2022}\natexlab{}.
\newblock \showarticletitle{ProcTHOR: Large-Scale Embodied AI Using Procedural Generation}.
\newblock \bibinfo{journal}{\emph{Advances in Neural Information Processing Systems}}  \bibinfo{volume}{35} (\bibinfo{year}{2022}), \bibinfo{pages}{5982--5994}.
\newblock


\bibitem[Dekel et~al\mbox{.}(2023)]%
        {dekel2023speak}
\bibfield{author}{\bibinfo{person}{Avihu Dekel}, \bibinfo{person}{Slava Shechtman}, \bibinfo{person}{Raul Fernandez}, \bibinfo{person}{David Haws}, \bibinfo{person}{Zvi Kons}, {and} \bibinfo{person}{Ron Hoory}.} \bibinfo{year}{2023}\natexlab{}.
\newblock \showarticletitle{Speak While You Think: Streaming Speech Synthesis During Text Generation}.
\newblock \bibinfo{journal}{\emph{arXiv preprint arXiv:2309.11210}} (\bibinfo{year}{2023}).
\newblock


\bibitem[Dubois et~al\mbox{.}(2023)]%
        {dubois2023alpacafarm}
\bibfield{author}{\bibinfo{person}{Yann Dubois}, \bibinfo{person}{Xuechen Li}, \bibinfo{person}{Rohan Taori}, \bibinfo{person}{Tianyi Zhang}, \bibinfo{person}{Ishaan Gulrajani}, \bibinfo{person}{Jimmy Ba}, \bibinfo{person}{Carlos Guestrin}, \bibinfo{person}{Percy Liang}, {and} \bibinfo{person}{Tatsunori~B Hashimoto}.} \bibinfo{year}{2023}\natexlab{}.
\newblock \showarticletitle{Alpacafarm: A simulation framework for methods that learn from human feedback}.
\newblock \bibinfo{journal}{\emph{arXiv preprint arXiv:2305.14387}} (\bibinfo{year}{2023}).
\newblock


\bibitem[Duong and Solomon(2023)]%
        {duong2023analysis}
\bibfield{author}{\bibinfo{person}{Dat Duong} {and} \bibinfo{person}{Benjamin~D Solomon}.} \bibinfo{year}{2023}\natexlab{}.
\newblock \showarticletitle{Analysis of large-language model versus human performance for genetics questions}.
\newblock \bibinfo{journal}{\emph{European Journal of Human Genetics}} (\bibinfo{year}{2023}), \bibinfo{pages}{1--3}.
\newblock


\bibitem[Frank(2023)]%
        {frank2023baby}
\bibfield{author}{\bibinfo{person}{Michael~C Frank}.} \bibinfo{year}{2023}\natexlab{}.
\newblock \showarticletitle{Baby steps in evaluating the capacities of large language models}.
\newblock \bibinfo{journal}{\emph{Nature Reviews Psychology}} \bibinfo{volume}{2}, \bibinfo{number}{8} (\bibinfo{year}{2023}), \bibinfo{pages}{451--452}.
\newblock


\bibitem[Gal et~al\mbox{.}(2023)]%
        {gal2023encoder}
\bibfield{author}{\bibinfo{person}{Rinon Gal}, \bibinfo{person}{Moab Arar}, \bibinfo{person}{Yuval Atzmon}, \bibinfo{person}{Amit~H Bermano}, \bibinfo{person}{Gal Chechik}, {and} \bibinfo{person}{Daniel Cohen-Or}.} \bibinfo{year}{2023}\natexlab{}.
\newblock \showarticletitle{Encoder-based domain tuning for fast personalization of text-to-image models}.
\newblock \bibinfo{journal}{\emph{ACM Transactions on Graphics (TOG)}} (\bibinfo{year}{2023}).
\newblock


\bibitem[Gehman et~al\mbox{.}(2020)]%
        {gehman2020realtoxicityprompts}
\bibfield{author}{\bibinfo{person}{Samuel Gehman}, \bibinfo{person}{Suchin Gururangan}, \bibinfo{person}{Maarten Sap}, \bibinfo{person}{Yejin Choi}, {and} \bibinfo{person}{Noah~A Smith}.} \bibinfo{year}{2020}\natexlab{}.
\newblock \showarticletitle{Realtoxicityprompts: Evaluating neural toxic degeneration in language models}.
\newblock \bibinfo{journal}{\emph{arXiv preprint arXiv:2009.11462}} (\bibinfo{year}{2020}).
\newblock


\bibitem[Helm et~al\mbox{.}(2023)]%
        {helm2023diversity}
\bibfield{author}{\bibinfo{person}{Paula Helm}, \bibinfo{person}{G{\'a}bor Bella}, \bibinfo{person}{Gertraud Koch}, {and} \bibinfo{person}{Fausto Giunchiglia}.} \bibinfo{year}{2023}\natexlab{}.
\newblock \showarticletitle{Diversity and language technology: how techno-linguistic bias can cause epistemic injustice}.
\newblock \bibinfo{journal}{\emph{arXiv preprint arXiv:2307.13714}} (\bibinfo{year}{2023}).
\newblock


\bibitem[Hendrycks et~al\mbox{.}(2020)]%
        {hendrycks2020measuring}
\bibfield{author}{\bibinfo{person}{Dan Hendrycks}, \bibinfo{person}{Collin Burns}, \bibinfo{person}{Steven Basart}, \bibinfo{person}{Andy Zou}, \bibinfo{person}{Mantas Mazeika}, \bibinfo{person}{Dawn Song}, {and} \bibinfo{person}{Jacob Steinhardt}.} \bibinfo{year}{2020}\natexlab{}.
\newblock \showarticletitle{Measuring massive multitask language understanding}.
\newblock \bibinfo{journal}{\emph{arXiv preprint arXiv:2009.03300}} (\bibinfo{year}{2020}).
\newblock


\bibitem[Huang et~al\mbox{.}(2023)]%
        {huang2023language}
\bibfield{author}{\bibinfo{person}{Shaohan Huang}, \bibinfo{person}{Li Dong}, \bibinfo{person}{Wenhui Wang}, \bibinfo{person}{Yaru Hao}, \bibinfo{person}{Saksham Singhal}, \bibinfo{person}{Shuming Ma}, \bibinfo{person}{Tengchao Lv}, \bibinfo{person}{Lei Cui}, \bibinfo{person}{Owais~Khan Mohammed}, \bibinfo{person}{Qiang Liu}, {et~al\mbox{.}}} \bibinfo{year}{2023}\natexlab{}.
\newblock \showarticletitle{Language is not all you need: Aligning perception with language models}.
\newblock \bibinfo{journal}{\emph{arXiv preprint arXiv:2302.14045}} (\bibinfo{year}{2023}).
\newblock


\bibitem[Jansen(2006)]%
        {jansen2006search}
\bibfield{author}{\bibinfo{person}{Bernard~J Jansen}.} \bibinfo{year}{2006}\natexlab{}.
\newblock \showarticletitle{Search log analysis: What it is, what's been done, how to do it}.
\newblock \bibinfo{journal}{\emph{Library \& information science research}} \bibinfo{volume}{28}, \bibinfo{number}{3} (\bibinfo{year}{2006}), \bibinfo{pages}{407--432}.
\newblock


\bibitem[Jansen et~al\mbox{.}(2007)]%
        {jansen2007determining}
\bibfield{author}{\bibinfo{person}{Bernard~J Jansen}, \bibinfo{person}{Danielle~L Booth}, {and} \bibinfo{person}{Amanda Spink}.} \bibinfo{year}{2007}\natexlab{}.
\newblock \showarticletitle{Determining the user intent of web search engine queries}. In \bibinfo{booktitle}{\emph{Proceedings of the 16th international conference on World Wide Web}}. \bibinfo{pages}{1149--1150}.
\newblock


\bibitem[Jiang et~al\mbox{.}(2023)]%
        {jiang2023mistral}
\bibfield{author}{\bibinfo{person}{Albert~Q Jiang}, \bibinfo{person}{Alexandre Sablayrolles}, \bibinfo{person}{Arthur Mensch}, \bibinfo{person}{Chris Bamford}, \bibinfo{person}{Devendra~Singh Chaplot}, \bibinfo{person}{Diego de~las Casas}, \bibinfo{person}{Florian Bressand}, \bibinfo{person}{Gianna Lengyel}, \bibinfo{person}{Guillaume Lample}, \bibinfo{person}{Lucile Saulnier}, {et~al\mbox{.}}} \bibinfo{year}{2023}\natexlab{}.
\newblock \showarticletitle{Mistral 7B}.
\newblock \bibinfo{journal}{\emph{arXiv preprint arXiv:2310.06825}} (\bibinfo{year}{2023}).
\newblock


\bibitem[Kaufmann et~al\mbox{.}(2023)]%
        {kaufmann2023survey}
\bibfield{author}{\bibinfo{person}{Timo Kaufmann}, \bibinfo{person}{Paul Weng}, \bibinfo{person}{Viktor Bengs}, {and} \bibinfo{person}{Eyke H{\"u}llermeier}.} \bibinfo{year}{2023}\natexlab{}.
\newblock \showarticletitle{A Survey of Reinforcement Learning from Human Feedback}.
\newblock \bibinfo{journal}{\emph{arXiv preprint arXiv:2312.14925}} (\bibinfo{year}{2023}).
\newblock


\bibitem[Kelly et~al\mbox{.}(2023)]%
        {kelly2023bing}
\bibfield{author}{\bibinfo{person}{Dominique Kelly}, \bibinfo{person}{Yimin Chen}, \bibinfo{person}{Sarah~E Cornwell}, \bibinfo{person}{Nicole~S Delellis}, \bibinfo{person}{Alex Mayhew}, \bibinfo{person}{Sodiq Onaolapo}, {and} \bibinfo{person}{Victoria~L Rubin}.} \bibinfo{year}{2023}\natexlab{}.
\newblock \showarticletitle{Bing Chat: The Future of Search Engines?}
\newblock \bibinfo{journal}{\emph{Proceedings of the Association for Information Science and Technology}} (\bibinfo{year}{2023}).
\newblock


\bibitem[Kofler et~al\mbox{.}(2016)]%
        {kofler2016user}
\bibfield{author}{\bibinfo{person}{Christoph Kofler}, \bibinfo{person}{Martha Larson}, {and} \bibinfo{person}{Alan Hanjalic}.} \bibinfo{year}{2016}\natexlab{}.
\newblock \showarticletitle{User intent in multimedia search: a survey of the state of the art and future challenges}.
\newblock \bibinfo{journal}{\emph{ACM Computing Surveys (CSUR)}} \bibinfo{volume}{49}, \bibinfo{number}{2} (\bibinfo{year}{2016}), \bibinfo{pages}{1--37}.
\newblock


\bibitem[Lai et~al\mbox{.}(2021)]%
        {lai2021towards}
\bibfield{author}{\bibinfo{person}{Vivian Lai}, \bibinfo{person}{Chacha Chen}, \bibinfo{person}{Q~Vera Liao}, \bibinfo{person}{Alison Smith-Renner}, {and} \bibinfo{person}{Chenhao Tan}.} \bibinfo{year}{2021}\natexlab{}.
\newblock \showarticletitle{Towards a science of human-ai decision making: a survey of empirical studies}.
\newblock \bibinfo{journal}{\emph{arXiv preprint arXiv:2112.11471}} (\bibinfo{year}{2021}).
\newblock


\bibitem[Lambert and Calandra(2023)]%
        {lambert2023alignment}
\bibfield{author}{\bibinfo{person}{Nathan Lambert} {and} \bibinfo{person}{Roberto Calandra}.} \bibinfo{year}{2023}\natexlab{}.
\newblock \showarticletitle{The Alignment Ceiling: Objective Mismatch in Reinforcement Learning from Human Feedback}.
\newblock \bibinfo{journal}{\emph{arXiv preprint arXiv:2311.00168}} (\bibinfo{year}{2023}).
\newblock


\bibitem[Lanzi and Loiacono(2023)]%
        {lanzi2023chatgpt}
\bibfield{author}{\bibinfo{person}{Pier~Luca Lanzi} {and} \bibinfo{person}{Daniele Loiacono}.} \bibinfo{year}{2023}\natexlab{}.
\newblock \showarticletitle{Chatgpt and other large language models as evolutionary engines for online interactive collaborative game design}.
\newblock \bibinfo{journal}{\emph{arXiv preprint arXiv:2303.02155}} (\bibinfo{year}{2023}).
\newblock


\bibitem[Li et~al\mbox{.}(2023)]%
        {li2023alpacaeval}
\bibfield{author}{\bibinfo{person}{Xuechen Li}, \bibinfo{person}{Tianyi Zhang}, \bibinfo{person}{Yann Dubois}, \bibinfo{person}{Rohan Taori}, \bibinfo{person}{Ishaan Gulrajani}, \bibinfo{person}{Carlos Guestrin}, \bibinfo{person}{Percy Liang}, {and} \bibinfo{person}{Tatsunori~B Hashimoto}.} \bibinfo{year}{2023}\natexlab{}.
\newblock \showarticletitle{Alpacaeval: An automatic evaluator of instruction-following models}.
\newblock \bibinfo{journal}{\emph{GitHub repository}} (\bibinfo{year}{2023}).
\newblock


\bibitem[Liang et~al\mbox{.}(2022)]%
        {liang2022holistic}
\bibfield{author}{\bibinfo{person}{Percy Liang}, \bibinfo{person}{Rishi Bommasani}, \bibinfo{person}{Tony Lee}, \bibinfo{person}{Dimitris Tsipras}, \bibinfo{person}{Dilara Soylu}, \bibinfo{person}{Michihiro Yasunaga}, \bibinfo{person}{Yian Zhang}, \bibinfo{person}{Deepak Narayanan}, \bibinfo{person}{Yuhuai Wu}, \bibinfo{person}{Ananya Kumar}, {et~al\mbox{.}}} \bibinfo{year}{2022}\natexlab{}.
\newblock \showarticletitle{Holistic evaluation of language models}.
\newblock \bibinfo{journal}{\emph{arXiv preprint}} (\bibinfo{year}{2022}).
\newblock


\bibitem[Lin et~al\mbox{.}(2020)]%
        {lin2020limitations}
\bibfield{author}{\bibinfo{person}{Chu-Cheng Lin}, \bibinfo{person}{Aaron Jaech}, \bibinfo{person}{Xin Li}, \bibinfo{person}{Matthew~R Gormley}, {and} \bibinfo{person}{Jason Eisner}.} \bibinfo{year}{2020}\natexlab{}.
\newblock \showarticletitle{Limitations of autoregressive models and their alternatives}.
\newblock \bibinfo{journal}{\emph{arXiv preprint arXiv:2010.11939}} (\bibinfo{year}{2020}).
\newblock


\bibitem[Liu et~al\mbox{.}(2023)]%
        {liu2023trustworthy}
\bibfield{author}{\bibinfo{person}{Yang Liu}, \bibinfo{person}{Yuanshun Yao}, \bibinfo{person}{Jean-Francois Ton}, \bibinfo{person}{Xiaoying Zhang}, \bibinfo{person}{Ruocheng Guo~Hao Cheng}, \bibinfo{person}{Yegor Klochkov}, \bibinfo{person}{Muhammad~Faaiz Taufiq}, {and} \bibinfo{person}{Hang Li}.} \bibinfo{year}{2023}\natexlab{}.
\newblock \showarticletitle{Trustworthy LLMs: a Survey and Guideline for Evaluating Large Language Models' Alignment}.
\newblock \bibinfo{journal}{\emph{arXiv preprint arXiv:2308.05374}} (\bibinfo{year}{2023}).
\newblock


\bibitem[Lu et~al\mbox{.}(2015)]%
        {lu2015recommender}
\bibfield{author}{\bibinfo{person}{Jie Lu}, \bibinfo{person}{Dianshuang Wu}, \bibinfo{person}{Mingsong Mao}, \bibinfo{person}{Wei Wang}, {and} \bibinfo{person}{Guangquan Zhang}.} \bibinfo{year}{2015}\natexlab{}.
\newblock \showarticletitle{Recommender system application developments: a survey}.
\newblock \bibinfo{journal}{\emph{Decision support systems}}  \bibinfo{volume}{74} (\bibinfo{year}{2015}), \bibinfo{pages}{12--32}.
\newblock


\bibitem[Madasu and Srivastava(2022)]%
        {madasu2022large}
\bibfield{author}{\bibinfo{person}{Avinash Madasu} {and} \bibinfo{person}{Shashank Srivastava}.} \bibinfo{year}{2022}\natexlab{}.
\newblock \showarticletitle{What do Large Language Models Learn beyond Language?}
\newblock \bibinfo{journal}{\emph{arXiv preprint arXiv:2210.12302}} (\bibinfo{year}{2022}).
\newblock


\bibitem[Mikalef and Gupta(2021)]%
        {mikalef2021artificial}
\bibfield{author}{\bibinfo{person}{Patrick Mikalef} {and} \bibinfo{person}{Manjul Gupta}.} \bibinfo{year}{2021}\natexlab{}.
\newblock \showarticletitle{Artificial intelligence capability: Conceptualization, measurement calibration, and empirical study on its impact on organizational creativity and firm performance}.
\newblock \bibinfo{journal}{\emph{Information \& Management}} (\bibinfo{year}{2021}).
\newblock


\bibitem[Paulik et~al\mbox{.}(2021)]%
        {paulik2021federated}
\bibfield{author}{\bibinfo{person}{Matthias Paulik}, \bibinfo{person}{Matt Seigel}, \bibinfo{person}{Henry Mason}, \bibinfo{person}{Dominic Telaar}, \bibinfo{person}{Joris Kluivers}, \bibinfo{person}{Rogier van Dalen}, \bibinfo{person}{Chi~Wai Lau}, \bibinfo{person}{Luke Carlson}, \bibinfo{person}{Filip Granqvist}, \bibinfo{person}{Chris Vandevelde}, {et~al\mbox{.}}} \bibinfo{year}{2021}\natexlab{}.
\newblock \showarticletitle{Federated evaluation and tuning for on-device personalization: System design \& applications}.
\newblock \bibinfo{journal}{\emph{arXiv preprint arXiv:2102.08503}} (\bibinfo{year}{2021}).
\newblock


\bibitem[Qin et~al\mbox{.}(2023)]%
        {qin2023tool}
\bibfield{author}{\bibinfo{person}{Yujia Qin}, \bibinfo{person}{Shengding Hu}, \bibinfo{person}{Yankai Lin}, \bibinfo{person}{Weize Chen}, \bibinfo{person}{Ning Ding}, \bibinfo{person}{Ganqu Cui}, \bibinfo{person}{Zheni Zeng}, \bibinfo{person}{Yufei Huang}, \bibinfo{person}{Chaojun Xiao}, \bibinfo{person}{Chi Han}, {et~al\mbox{.}}} \bibinfo{year}{2023}\natexlab{}.
\newblock \showarticletitle{Tool learning with foundation models}.
\newblock \bibinfo{journal}{\emph{arXiv preprint arXiv:2304.08354}} (\bibinfo{year}{2023}).
\newblock


\bibitem[Qu et~al\mbox{.}(2018)]%
        {qu2018analyzing}
\bibfield{author}{\bibinfo{person}{Chen Qu}, \bibinfo{person}{Liu Yang}, \bibinfo{person}{W~Bruce Croft}, \bibinfo{person}{Johanne~R Trippas}, \bibinfo{person}{Yongfeng Zhang}, {and} \bibinfo{person}{Minghui Qiu}.} \bibinfo{year}{2018}\natexlab{}.
\newblock \showarticletitle{Analyzing and characterizing user intent in information-seeking conversations}. In \bibinfo{booktitle}{\emph{The 41st international acm sigir conference on research \& development in information retrieval}}. \bibinfo{pages}{989--992}.
\newblock


\bibitem[Qu et~al\mbox{.}(2019)]%
        {qu2019user}
\bibfield{author}{\bibinfo{person}{Chen Qu}, \bibinfo{person}{Liu Yang}, \bibinfo{person}{W~Bruce Croft}, \bibinfo{person}{Yongfeng Zhang}, \bibinfo{person}{Johanne~R Trippas}, {and} \bibinfo{person}{Minghui Qiu}.} \bibinfo{year}{2019}\natexlab{}.
\newblock \showarticletitle{User intent prediction in information-seeking conversations}. In \bibinfo{booktitle}{\emph{Proceedings of the 2019 Conference on Human Information Interaction and Retrieval}}. \bibinfo{pages}{25--33}.
\newblock


\bibitem[Rahman et~al\mbox{.}(2023)]%
        {rahman2023quantized}
\bibfield{author}{\bibinfo{person}{Mohammad Wali~Ur Rahman}, \bibinfo{person}{Murad~Mehrab Abrar}, \bibinfo{person}{Hunter~Gibbons Copening}, \bibinfo{person}{Salim Hariri}, \bibinfo{person}{Sicong Shao}, \bibinfo{person}{Pratik Satam}, {and} \bibinfo{person}{Soheil Salehi}.} \bibinfo{year}{2023}\natexlab{}.
\newblock \showarticletitle{Quantized Transformer Language Model Implementations on Edge Devices}.
\newblock \bibinfo{journal}{\emph{arXiv preprint arXiv:2310.03971}} (\bibinfo{year}{2023}).
\newblock


\bibitem[Sakaguchi et~al\mbox{.}(2021)]%
        {sakaguchi2021winogrande}
\bibfield{author}{\bibinfo{person}{Keisuke Sakaguchi}, \bibinfo{person}{Ronan~Le Bras}, \bibinfo{person}{Chandra Bhagavatula}, {and} \bibinfo{person}{Yejin Choi}.} \bibinfo{year}{2021}\natexlab{}.
\newblock \showarticletitle{Winogrande: An adversarial winograd schema challenge at scale}.
\newblock \bibinfo{journal}{\emph{Commun. ACM}} \bibinfo{volume}{64}, \bibinfo{number}{9} (\bibinfo{year}{2021}), \bibinfo{pages}{99--106}.
\newblock


\bibitem[Saydam et~al\mbox{.}(2022)]%
        {saydam2022does}
\bibfield{author}{\bibinfo{person}{Mehmet~Bahri Saydam}, \bibinfo{person}{Hasan~Evrim Arici}, {and} \bibinfo{person}{Mehmet~Ali Koseoglu}.} \bibinfo{year}{2022}\natexlab{}.
\newblock \showarticletitle{How does the tourism and hospitality industry use artificial intelligence? A review of empirical studies and future research agenda}.
\newblock \bibinfo{journal}{\emph{Journal of Hospitality Marketing \& Management}} (\bibinfo{year}{2022}).
\newblock


\bibitem[Shah and Bender(2022)]%
        {shah2022situating}
\bibfield{author}{\bibinfo{person}{Chirag Shah} {and} \bibinfo{person}{Emily~M Bender}.} \bibinfo{year}{2022}\natexlab{}.
\newblock \showarticletitle{Situating search}. In \bibinfo{booktitle}{\emph{Proceedings of the 2022 Conference on Human Information Interaction and Retrieval}}. \bibinfo{pages}{221--232}.
\newblock


\bibitem[Shah et~al\mbox{.}(2023)]%
        {shah2023using}
\bibfield{author}{\bibinfo{person}{Chirag Shah}, \bibinfo{person}{Ryen~W White}, \bibinfo{person}{Reid Andersen}, \bibinfo{person}{Georg Buscher}, \bibinfo{person}{Scott Counts}, \bibinfo{person}{Sarkar Snigdha~Sarathi Das}, \bibinfo{person}{Ali Montazer}, \bibinfo{person}{Sathish Manivannan}, \bibinfo{person}{Jennifer Neville}, \bibinfo{person}{Xiaochuan Ni}, {et~al\mbox{.}}} \bibinfo{year}{2023}\natexlab{}.
\newblock \showarticletitle{Using large language models to generate, validate, and apply user intent taxonomies}.
\newblock \bibinfo{journal}{\emph{arXiv preprint arXiv:2309.13063}} (\bibinfo{year}{2023}).
\newblock


\bibitem[Shanahan and Clarke(2023)]%
        {shanahan2023evaluating}
\bibfield{author}{\bibinfo{person}{Murray Shanahan} {and} \bibinfo{person}{Catherine Clarke}.} \bibinfo{year}{2023}\natexlab{}.
\newblock \showarticletitle{Evaluating Large Language Model Creativity from a Literary Perspective}.
\newblock \bibinfo{journal}{\emph{arXiv preprint arXiv:2312.03746}} (\bibinfo{year}{2023}).
\newblock


\bibitem[Sinha et~al\mbox{.}(2023)]%
        {sinha2023mathematical}
\bibfield{author}{\bibinfo{person}{Ritwik Sinha}, \bibinfo{person}{Zhao Song}, {and} \bibinfo{person}{Tianyi Zhou}.} \bibinfo{year}{2023}\natexlab{}.
\newblock \showarticletitle{A mathematical abstraction for balancing the trade-off between creativity and reality in large language models}.
\newblock \bibinfo{journal}{\emph{arXiv preprint arXiv:2306.02295}} (\bibinfo{year}{2023}).
\newblock


\bibitem[Song et~al\mbox{.}(2022)]%
        {song2022can}
\bibfield{author}{\bibinfo{person}{Xia Song}, \bibinfo{person}{Bo Xu}, {and} \bibinfo{person}{Zhenzhen Zhao}.} \bibinfo{year}{2022}\natexlab{}.
\newblock \showarticletitle{Can people experience romantic love for artificial intelligence? An empirical study of intelligent assistants}.
\newblock \bibinfo{journal}{\emph{Information \& Management}} (\bibinfo{year}{2022}).
\newblock


\bibitem[Su et~al\mbox{.}(2018)]%
        {su2018user}
\bibfield{author}{\bibinfo{person}{Ning Su}, \bibinfo{person}{Jiyin He}, \bibinfo{person}{Yiqun Liu}, \bibinfo{person}{Min Zhang}, {and} \bibinfo{person}{Shaoping Ma}.} \bibinfo{year}{2018}\natexlab{}.
\newblock \showarticletitle{User intent, behaviour, and perceived satisfaction in product search}. In \bibinfo{booktitle}{\emph{Proceedings of the Eleventh ACM International Conference on Web Search and Data Mining}}. \bibinfo{pages}{547--555}.
\newblock


\bibitem[Suzgun et~al\mbox{.}(2022)]%
        {suzgun2022challenging}
\bibfield{author}{\bibinfo{person}{Mirac Suzgun}, \bibinfo{person}{Nathan Scales}, \bibinfo{person}{Nathanael Sch{\"a}rli}, \bibinfo{person}{Sebastian Gehrmann}, \bibinfo{person}{Yi Tay}, \bibinfo{person}{Hyung~Won Chung}, \bibinfo{person}{Aakanksha Chowdhery}, \bibinfo{person}{Quoc~V Le}, \bibinfo{person}{Ed~H Chi}, \bibinfo{person}{Denny Zhou}, {et~al\mbox{.}}} \bibinfo{year}{2022}\natexlab{}.
\newblock \showarticletitle{Challenging big-bench tasks and whether chain-of-thought can solve them}.
\newblock \bibinfo{journal}{\emph{arXiv preprint arXiv:2210.09261}} (\bibinfo{year}{2022}).
\newblock


\bibitem[Tang et~al\mbox{.}(2023)]%
        {tang2023medical}
\bibfield{author}{\bibinfo{person}{Lu Tang}, \bibinfo{person}{Jinxu Li}, {and} \bibinfo{person}{Sophia Fantus}.} \bibinfo{year}{2023}\natexlab{}.
\newblock \showarticletitle{Medical artificial intelligence ethics: A systematic review of empirical studies}.
\newblock \bibinfo{journal}{\emph{Digital Health}} (\bibinfo{year}{2023}).
\newblock


\bibitem[Tang et~al\mbox{.}(2011)]%
        {tang2011intentsearch}
\bibfield{author}{\bibinfo{person}{Xiaoou Tang}, \bibinfo{person}{Ke Liu}, \bibinfo{person}{Jingyu Cui}, \bibinfo{person}{Fang Wen}, {and} \bibinfo{person}{Xiaogang Wang}.} \bibinfo{year}{2011}\natexlab{}.
\newblock \showarticletitle{Intentsearch: Capturing user intention for one-click internet image search}.
\newblock \bibinfo{journal}{\emph{IEEE transactions on pattern analysis and machine intelligence}} \bibinfo{volume}{34}, \bibinfo{number}{7} (\bibinfo{year}{2011}), \bibinfo{pages}{1342--1353}.
\newblock


\bibitem[Team et~al\mbox{.}(2023)]%
        {team2023gemini}
\bibfield{author}{\bibinfo{person}{Gemini Team}, \bibinfo{person}{Rohan Anil}, \bibinfo{person}{Sebastian Borgeaud}, \bibinfo{person}{Yonghui Wu}, \bibinfo{person}{Jean-Baptiste Alayrac}, \bibinfo{person}{Jiahui Yu}, \bibinfo{person}{Radu Soricut}, \bibinfo{person}{Johan Schalkwyk}, \bibinfo{person}{Andrew~M Dai}, \bibinfo{person}{Anja Hauth}, {et~al\mbox{.}}} \bibinfo{year}{2023}\natexlab{}.
\newblock \showarticletitle{Gemini: a family of highly capable multimodal models}.
\newblock \bibinfo{journal}{\emph{arXiv preprint arXiv:2312.11805}} (\bibinfo{year}{2023}).
\newblock


\bibitem[Teevan et~al\mbox{.}(2008)]%
        {teevan2008personalize}
\bibfield{author}{\bibinfo{person}{Jaime Teevan}, \bibinfo{person}{Susan~T Dumais}, {and} \bibinfo{person}{Daniel~J Liebling}.} \bibinfo{year}{2008}\natexlab{}.
\newblock \showarticletitle{To personalize or not to personalize: modeling queries with variation in user intent}. In \bibinfo{booktitle}{\emph{Proceedings of the 31st annual international ACM SIGIR conference}}.
\newblock


\bibitem[Touvron et~al\mbox{.}(2023)]%
        {touvron2023llama}
\bibfield{author}{\bibinfo{person}{Hugo Touvron}, \bibinfo{person}{Louis Martin}, \bibinfo{person}{Kevin Stone}, \bibinfo{person}{Peter Albert}, \bibinfo{person}{Amjad Almahairi}, \bibinfo{person}{Yasmine Babaei}, \bibinfo{person}{Nikolay Bashlykov}, \bibinfo{person}{Soumya Batra}, \bibinfo{person}{Prajjwal Bhargava}, \bibinfo{person}{Shruti Bhosale}, {et~al\mbox{.}}} \bibinfo{year}{2023}\natexlab{}.
\newblock \showarticletitle{Llama 2: Open foundation and fine-tuned chat models, 2023}.
\newblock \bibinfo{journal}{\emph{URL https://arxiv. org/abs/2307.09288}} (\bibinfo{year}{2023}).
\newblock


\bibitem[Vereschak et~al\mbox{.}(2021)]%
        {vereschak2021evaluate}
\bibfield{author}{\bibinfo{person}{Oleksandra Vereschak}, \bibinfo{person}{Gilles Bailly}, {and} \bibinfo{person}{Baptiste Caramiaux}.} \bibinfo{year}{2021}\natexlab{}.
\newblock \showarticletitle{How to evaluate trust in AI-assisted decision making? A survey of empirical methodologies}.
\newblock \bibinfo{journal}{\emph{Proceedings of the ACM on Human-Computer Interaction}} (\bibinfo{year}{2021}).
\newblock


\bibitem[Wang et~al\mbox{.}(2023)]%
        {wang2023robustness}
\bibfield{author}{\bibinfo{person}{Jindong Wang}, \bibinfo{person}{Xixu Hu}, \bibinfo{person}{Wenxin Hou}, \bibinfo{person}{Hao Chen}, \bibinfo{person}{Runkai Zheng}, \bibinfo{person}{Yidong Wang}, \bibinfo{person}{Linyi Yang}, \bibinfo{person}{Haojun Huang}, \bibinfo{person}{Wei Ye}, \bibinfo{person}{Xiubo Geng}, {et~al\mbox{.}}} \bibinfo{year}{2023}\natexlab{}.
\newblock \showarticletitle{On the robustness of chatgpt: An adversarial and out-of-distribution perspective}.
\newblock \bibinfo{journal}{\emph{arXiv preprint arXiv:2302.12095}} (\bibinfo{year}{2023}).
\newblock


\bibitem[Xia et~al\mbox{.}(2018)]%
        {xia2018zero}
\bibfield{author}{\bibinfo{person}{Congying Xia}, \bibinfo{person}{Chenwei Zhang}, \bibinfo{person}{Xiaohui Yan}, \bibinfo{person}{Yi Chang}, {and} \bibinfo{person}{Philip~S Yu}.} \bibinfo{year}{2018}\natexlab{}.
\newblock \showarticletitle{Zero-shot user intent detection via capsule neural networks}.
\newblock \bibinfo{journal}{\emph{arXiv preprint arXiv:1809.00385}} (\bibinfo{year}{2018}).
\newblock


\bibitem[Xiao et~al\mbox{.}(2023b)]%
        {xiao2023plug}
\bibfield{author}{\bibinfo{person}{Chaojun Xiao}, \bibinfo{person}{Zhengyan Zhang}, \bibinfo{person}{Xu Han}, \bibinfo{person}{Chi-Min Chan}, \bibinfo{person}{Yankai Lin}, \bibinfo{person}{Zhiyuan Liu}, \bibinfo{person}{Xiangyang Li}, \bibinfo{person}{Zhonghua Li}, \bibinfo{person}{Zhao Cao}, {and} \bibinfo{person}{Maosong Sun}.} \bibinfo{year}{2023}\natexlab{b}.
\newblock \showarticletitle{Plug-and-play document modules for pre-trained models}.
\newblock  (\bibinfo{year}{2023}).
\newblock


\bibitem[Xiao et~al\mbox{.}(2023a)]%
        {xiao2023efficient}
\bibfield{author}{\bibinfo{person}{Guangxuan Xiao}, \bibinfo{person}{Yuandong Tian}, \bibinfo{person}{Beidi Chen}, \bibinfo{person}{Song Han}, {and} \bibinfo{person}{Mike Lewis}.} \bibinfo{year}{2023}\natexlab{a}.
\newblock \showarticletitle{Efficient streaming language models with attention sinks}.
\newblock \bibinfo{journal}{\emph{arXiv preprint arXiv:2309.17453}} (\bibinfo{year}{2023}).
\newblock


\bibitem[Xie et~al\mbox{.}(2023)]%
        {xie2023prompt}
\bibfield{author}{\bibinfo{person}{Yutong Xie}, \bibinfo{person}{Zhaoying Pan}, \bibinfo{person}{Jinge Ma}, \bibinfo{person}{Luo Jie}, {and} \bibinfo{person}{Qiaozhu Mei}.} \bibinfo{year}{2023}\natexlab{}.
\newblock \showarticletitle{A prompt log analysis of text-to-image generation systems}. In \bibinfo{booktitle}{\emph{Proceedings of the ACM Web Conference 2023}}. \bibinfo{pages}{3892--3902}.
\newblock


\bibitem[Yang et~al\mbox{.}(2023)]%
        {yang2023holodeck}
\bibfield{author}{\bibinfo{person}{Yue Yang}, \bibinfo{person}{Fan-Yun Sun}, \bibinfo{person}{Luca Weihs}, \bibinfo{person}{Eli VanderBilt}, \bibinfo{person}{Alvaro Herrasti}, \bibinfo{person}{Winson Han}, \bibinfo{person}{Jiajun Wu}, \bibinfo{person}{Nick Haber}, \bibinfo{person}{Ranjay Krishna}, \bibinfo{person}{Lingjie Liu}, {et~al\mbox{.}}} \bibinfo{year}{2023}\natexlab{}.
\newblock \showarticletitle{Holodeck: Language Guided Generation of 3D Embodied AI Environments}.
\newblock \bibinfo{journal}{\emph{arXiv preprint arXiv:2312.09067}} (\bibinfo{year}{2023}).
\newblock


\bibitem[Yi et~al\mbox{.}(2009)]%
        {yi2009discovering}
\bibfield{author}{\bibinfo{person}{Xing Yi}, \bibinfo{person}{Hema Raghavan}, {and} \bibinfo{person}{Chris Leggetter}.} \bibinfo{year}{2009}\natexlab{}.
\newblock \showarticletitle{Discovering users' specific geo intention in web search}. In \bibinfo{booktitle}{\emph{Proceedings of the 18th international conference on World wide web}}. \bibinfo{pages}{481--490}.
\newblock


\bibitem[Zellers et~al\mbox{.}(2019)]%
        {zellers2019hellaswag}
\bibfield{author}{\bibinfo{person}{Rowan Zellers}, \bibinfo{person}{Ari Holtzman}, \bibinfo{person}{Yonatan Bisk}, \bibinfo{person}{Ali Farhadi}, {and} \bibinfo{person}{Yejin Choi}.} \bibinfo{year}{2019}\natexlab{}.
\newblock \showarticletitle{Hellaswag: Can a machine really finish your sentence?}
\newblock \bibinfo{journal}{\emph{arXiv preprint arXiv:1905.07830}} (\bibinfo{year}{2019}).
\newblock


\bibitem[Zhang et~al\mbox{.}(2023)]%
        {zhang2023plug}
\bibfield{author}{\bibinfo{person}{Zhengyan Zhang}, \bibinfo{person}{Zhiyuan Zeng}, \bibinfo{person}{Yankai Lin}, \bibinfo{person}{Huadong Wang}, \bibinfo{person}{Deming Ye}, \bibinfo{person}{Chaojun Xiao}, \bibinfo{person}{Xu Han}, \bibinfo{person}{Zhiyuan Liu}, \bibinfo{person}{Peng Li}, \bibinfo{person}{Maosong Sun}, {et~al\mbox{.}}} \bibinfo{year}{2023}\natexlab{}.
\newblock \showarticletitle{Plug-and-play knowledge injection for pre-trained language models}.
\newblock \bibinfo{journal}{\emph{arXiv preprint}} (\bibinfo{year}{2023}).
\newblock


\bibitem[Zhong et~al\mbox{.}(2023)]%
        {zhong2023agieval}
\bibfield{author}{\bibinfo{person}{Wanjun Zhong}, \bibinfo{person}{Ruixiang Cui}, \bibinfo{person}{Yiduo Guo}, \bibinfo{person}{Yaobo Liang}, \bibinfo{person}{Shuai Lu}, \bibinfo{person}{Yanlin Wang}, \bibinfo{person}{Amin Saied}, \bibinfo{person}{Weizhu Chen}, {and} \bibinfo{person}{Nan Duan}.} \bibinfo{year}{2023}\natexlab{}.
\newblock \showarticletitle{Agieval: A human-centric benchmark for evaluating foundation models}.
\newblock \bibinfo{journal}{\emph{arXiv preprint}} (\bibinfo{year}{2023}).
\newblock


\end{thebibliography}
